\documentclass[prd,showpacs,amsmath,amssymb,nofootinbib,superscriptaddress]{revtex4}

\usepackage{graphicx}% Include figure files
\usepackage{dcolumn}% Align table columns on decimal point
\usepackage{bm}% bold math
\usepackage{bbold}
\newcommand{\Od}{{\cal O}}
\newcommand{\diag}{\mbox{diag}}
\newcommand{\intT}{\int_0^\beta d\tau \int d^3 \vec{x}}
\newcommand{\intx}{\int d^4 x}
\newcommand{\mean}[1]{\left\langle{#1}\right\rangle}
\newcommand{\parent}[1]{\left({#1}\right)}
\newcommand{\condu}{\langle \bar u u \rangle}
\newcommand{\condd}{\langle \bar d d \rangle}
\newcommand{\conds}{\langle \bar s s \rangle}
\newcommand{\condl}{\mean{\bar q q}_l}
\newcommand{\cth}{c_\theta}
\newcommand{\sth}{s_\theta}
\newcommand{\mpi}{M_{0\pi}}
\newcommand{\mk}{M_{0K}}
\newcommand{\meta}{M_{0\eta}}
\newcommand{\metap}{M_{0\eta^\prime}}
\newcommand{\ID}{{\mathbb{1}}}

\begin{document}
%\baselineskip=20pt
% declarations for front matter

\title{Pseudoscalar susceptibilities and quark condensates: chiral restoration  and lattice screening masses.}
\author{A. G\'omez Nicola}
\email{gomez@fis.ucm.es}
\affiliation{Departamento de F\'{\i}sica
Te\'orica II. Univ. Complutense. 28040 Madrid. Spain.}
\author{J. Ruiz de Elvira}
\email{elvira@hiskp.uni-bonn.de}
%\affiliation{Departamento de F\'{\i}sica Te\'orica II. Univ.
%Complutense. 28040 Madrid. Spain.}
\affiliation{Helmholtz-Institut f\"ur Strahlen- und Kernphysik, Universit\"at Bonn, D-53115 Bonn, Germany}

\begin{abstract}
We derive the formal Ward identities relating pseudoscalar susceptibilities and quark condensates in three-flavor QCD, including consistently the $\eta$-$\eta'$ sector and  the $U_A(1)$ anomaly. 
These identities are verified in the low-energy realization provided by ChPT, both in the standard $SU(3)$ framework for the octet case and combining the use of the $U(3)$ framework and the large-$N_c$ expansion of QCD
to account properly for the nonet sector and anomalous  contributions. The analysis is performed including finite temperature corrections as well as  the calculation of $U(3)$ quark condensates and all pseudoscalar susceptibilities, which together with the full set of Ward identities, are new results of this work.  
Finally, the Ward identities are  used to derive scaling relations for pseudoscalar masses which explain the behavior with temperature of lattice screening masses near chiral symmetry restoration.
\end{abstract}

 \pacs{11.10.Wx, % Finite temperature field theory
 11.30.Rd, % Chiral symmetries
 12.39.Fe, % Chiral lagrangians
 12.38.Gc. % Lattice QCD calculations
  25.75.Nq. %Quark deconfinement, quark-gluon plasma production, and phase transitions
 25.75.-q % Relativistic heavy-ion collisions
 }
\maketitle

\section{Introduction}

The QCD transition involving chiral symmetry restoration and deconfinement plays a crucial role to understand the behavior of matter created in Relativistic Heavy Ion Collision experiments, such as those at RHIC and LHC. 
In this context, there have been very significant advances from lattice groups in the study of the phase diagram and other thermodynamic properties \cite{Aoki:2009sc,Borsanyi:2010bp,Bazavov:2011nk,Buchoff:2013nra,Bhattacharya:2014ara,Cossu:2015lnb}. From those analysis, there is a general agreement that chiral symmetry is restored in a crossover transition in the 2+1 flavor case for physical masses at zero baryon density, 
which becomes a second-order phase transition in the chiral limit of vanishing light quark masses. The transition temperature is about $T_c\sim 150 - 160$ MeV. 

It is important to provide as much theoretical support as possible to these lattice results. From the old days of the $O(4)$ model description of chiral symmetry restoration \cite{Pisarski:1983ms}, 
there has been a lot of progress in this area. On the one hand, the Hadron Resonance Gas approach describes effectively the system with all free states thermally available \cite{Huovinen:2009yb}, 
while effective chiral models including explicitly vector and axial-vector resonances have  been successful to explain properties such as the dilepton and photon spectra and the $\rho-a_1$ mixing/degeneration at the chiral transition \cite{Rapp:1999ej}. On the other hand, studies based on Chiral Perturbation Theory (ChPT) \cite{Gasser:1983yg,Gasser:1984gg} have been able to establish many relevant physical properties of the meson gas. 
In more detail, the pressure and the chiral restoring behavior of the quark condensate have been obtained up to NNLO \cite{Gasser:1986vb,Gerber:1988tt}, 
pion spectral properties have been studied in \cite{Schenk:1993ru} and unitarized interactions have allowed to describe properly thermal resonances and transport coefficients~\cite{FernandezFraile:2009mi}. 

In a recent analysis \cite{Nicola:2013vma}, we have shown that an operator Ward identity between the pseudoscalar susceptibility and the quark condensate allows to understand the scaling of lattice screening masses near $T_c$. Furthermore, the same identity, which we verified in two-flavor ChPT, allows to understand the behavior of chiral partners in the scalar-pseudoscalar sector through degeneration of the corresponding  susceptibilities, 
also  seen in lattice data \cite{Buchoff:2013nra,Bhattacharya:2014ara}. In that description, the $f_0(500)$ thermal state plays a crucial role, since it saturates the scalar susceptibility producing a peak compatible with the transition temperature observed in the lattice. 

In this work we will study the QCD formal Ward identities that arise between pseudoscalar susceptibilities and quark condensates for three flavors. 
These identities will be verified in the model-independent framework provided by SU(3) ChPT for the pion and kaon channels and U(3) ChPT for the $\eta-\eta'$ sector. 
The latter will actually require to  combine the expansion in low energies and temperatures with that in $1/N_c$,  the so-called $\delta$ expansion \cite{'tHooft:1973jz,Witten:1979kh,Rosenzweig:1979ay,Witten:1979vv,Coleman:1980mx,Veneziano:1980xs,DiVecchia:1980ve,Witten:1980sp,HerreraSiklody:1996pm,Kaiser:2000gs,Guo:2011pa,Guo:2012ym,Guo:2012yt,Guo:2015xva}, 
so that the nonet field, the $\eta'$, can be treated consistently on the same footing as the other eight pseudo-Goldstone bosons. In that way, we will extend our previous $SU(2)$ analysis \cite{Nicola:2013vma} in a very nontrivial way. On the one hand, we will provide the full $N_f=3$ set of Ward identities, which involve anomalous contributions.
On the other hand, we will generalize the identities obtained originally in~{\cite{Broadhurst:1974ng,Bochicchio:1985xa} in formal QCD and without including the $\eta-\eta'$ sector, i.e, only for the $SU(3)$ chiral group. 
Apart from the formal derivation, the explicit verification provided here of those identities in the low-energy and finite-temperature representation of the model-independent ChPT framework, 
helps to clarify the role of the nonet and of the axial anomaly. 
We remark that this type of identities have been assumed in the lattice~\cite{Bhattacharya:2014ara} and in other phenomenological works~\cite{Cui:2015xta}. 

An important part of the present work will consist in the extension to three flavors of the study of screening masses performed for two flavors in~\cite{Nicola:2013vma}. 
We will see that the Ward identities between pseudoscalar susceptibilities and quark condensates allow to understand the behavior with temperature of lattice screening masses in the pion, kaon and $\bar s s$ channels, 
connecting it with chiral symmetry restoration. 
 We will carry out a thorough analysis of lattice results in this context, paying special attention to the definition of subtracted quark condensate operators in the lattice which have the correct scaling and chiral restoration properties.  

The plan of the paper is as follows. In section~\ref{sec:identities} we will derive the relevant Ward identities for the different channels as arising formally from the QCD generating functional, 
paying special attention to the anomalous contributions in the $\eta-\eta'$ sector, as well as to the degenerate limit of equal quark masses. 
Then, we will compute and verify each identity in ChPT, calculating in turn all the involved  pseudoscalar susceptibilities in the sectors $\pi$, $K$  and $\eta-\eta'$,  with a particular emphasis on the latter sector,  for which the quark condensates  are also derived here for the first time. 
In section~\ref{sec:latt} we will provide a detailed analysis of the implications of these results on lattice screening masses and their behavior near $T_c$,  exploring  their correlation with chiral symmetry restoration precisely through these Ward identities. Finally, in section \ref{sec:conc} we will present our conclusions. Explicit ChPT results will be collected in  Appendix~\ref{sec:app}.

\section{Pseudo-scalar susceptibilities and quark-condensates: QCD Ward identities and their low-energy representation}
\label{sec:identities}

We will first proceed to the formal derivation of the relevant Ward identities from QCD. We follow similar steps as in \cite{Bochicchio:1985xa}, where analogous identities are considered for Wilson fermions in the degenerate chiral limit, i.e, equal quark masses and condensates, and for the members of the $SU(3)$ octet.  We start by writing the expected value of a local operator $\mathcal{O}(x_1,\cdots,x_n)$  from the QCD generating functional as:

\begin{equation}
\left\langle\mathcal{O}(x_1,\cdots,x_n)\right\rangle=Z^{-1}\int{[dG][d\bar\psi][d\psi]\mathcal{O}(x_1,\cdots,x_n)e^{S_{QCD}}},
\label{oexp}
\end{equation} 
where $G_\mu^a$, $\psi$ are gluon and quark fields respectively, $Z=\int{[dG][d\bar\psi][d\psi]e^{S_{QCD}}}$ is the partition function and $S_{QCD}=i\intx \mathcal{L}_{QCD}$ in Minkowski space-time, where the fermion QCD Lagrangian in the light quark sector is:

\begin{equation}\label{QCD}
\mathcal{L}_{QCD}=-\frac{1}{4}G_{\mu\nu}^aG^{\mu\nu}_a + \bar\psi \left(i\gamma^{\mu} D_\mu-\mathcal{M}\right)\psi,
\end{equation}
with $D_\mu=\partial_\mu+igG_\mu$, $G_\mu=G_\mu^a(\lambda_a/2)$, $g$ the QCD coupling constant, $G_{\mu\nu}^a=\partial_\mu G_\nu^a-\partial_\nu G_\mu^a-gf_{abc}G_\mu^b G_\nu^c$ and $\mathcal{M}=\mbox{diag}(m_u,m_d,m_s)$ the quark mass matrix.  

In Euclidean space-time at finite temperature $T$, we have $i\intx\rightarrow \intT\equiv\int_T dx$ with $\tau=ix_0$ and $(-,-,-,-)$ metric, $G_\tau=-iG_0$, $G_{\tau j}=-iG_{0j}$. We will start working in  Minkowski space-time,  performing the rotation to  the Euclidean one only for our final expressions so that they can be  evaluated at finite temperature.

We consider an infinitesimal local axial transformation on the fermion fields $\psi'=\psi+\delta\psi$, $\bar\psi'=\bar\psi+\delta\bar\psi$, where to $\Od(\alpha)$:
\begin{align}
\delta\psi(x)&=i\alpha_A^a(x)\frac{ \lambda_a}{2}\gamma_5\psi(x),\nonumber\\
 \delta\bar\psi(x)&=i\bar\psi(x)\alpha_A^a(x)\frac{ \lambda_a}{2}\gamma_5,
 \label{fermiontrans}
\end{align}
 with $\lambda^a/2$ the flavor group generators to be specified below. 

We can now write the expectation value given in \eqref{oexp} in terms of the transformed variables $\psi'$, $\bar\psi'$, so that linearly in $\alpha^a$ we get:
\begin{align}
%&=Z^{-1}\int{d[G]d[\psi]d[\bar\psi]\left\{\frac{\delta\mathcal{O}(x_1,\cdots,x_n)}{\delta \alpha_A^a(x)}+\mathcal{O}(x_1,\cdots,x_n)\left[\frac{\delta S}{\delta \alpha^a_A(x)}+\frac{\delta \log\mathcal{J}}{\delta \alpha_A^a (x)}\right]\right\}e^{S}}\nonumber\\
&\left\langle\frac{\delta\mathcal{O}(x_1,\cdots,x_n)}{\delta \alpha_A^a(x)}\right\rangle+\left\langle\mathcal{O}(x_1,\cdots,x_n)\frac{\delta S_{QCD}}{\delta \alpha_A^a(x)}\right\rangle 
+\left\langle \mathcal{O}(x_1,\cdots,x_n) \frac{\delta \log\mathcal{J}}{\delta \alpha_A^a (x)}\right\rangle=0,
\label{master}
\end{align}
where $\mathcal{J}$ is the anomalous jacobian of the fermionic measure under the axial transformation in \eqref{fermiontrans}, i.e $[d\bar \psi'][d\psi']=\mathcal{J} [d\bar \psi][d\psi]$. 

The second term in (\ref{master}) reads:
\begin{equation}\label{ST}
\left\langle\mathcal{O}(x_1,\cdots,x_n)\frac{\delta S_{QCD}}{\delta \alpha_A^a(x)}\right\rangle=i\partial_\mu^x\left\langle\mathcal{O}(x_1,\cdots,x_n)J_A^{a\mu}(x)\right\rangle+
\left\langle\mathcal{O}(x_1,\cdots,x_n)\left[\bar\psi(x)\{\frac{\lambda^a}{2},\mathcal{M}\}\gamma_5\psi(x)\right]\right\rangle,
\end{equation}
with the axial current $J_A^{a\mu}=\bar\psi\gamma^{\mu}\gamma_5 \frac{\lambda^a}{2}\psi$. 

A particular case of interest for our work is to evaluate \eqref{master} for a pseudo-scalar current, $\mathcal{O}(y)=P^b(y)=i\bar\psi(y)\gamma_5 \lambda^b\psi(y)$, which satisfies:

\begin{equation}\label{Pbtrans}
\frac{\delta P^b(y)}{\delta \alpha_A^a(x)}=-2\bar\psi(x)\left\{\frac{ \lambda^a}{2},\frac{ \lambda^b}{2}\right\}\psi(x)\delta^{(4)}(x-y).
\end{equation}

The generators $\lambda^a$ will be the Pauli matrices for the two-flavor case. For three flavors,  $\lambda^a$ will be the Gell-Mann matrices for $a=1,\dots,8$ (octet) and for the nonet we will consider in addition $\displaystyle\lambda^0=\sqrt{2/3}\,\ID$, which together with $\lambda^a$ expand the whole space of $3\times 3$ unitary matrices.  

Thus, for $a=1,\dots,8$, the transformation \eqref{fermiontrans} is not anomalous, so that $\log\mathcal{J}=0$, while for $a=0$, we have \cite{Fujikawa:1980eg}:

\begin{equation}
\log\mathcal{J}=-i\intx \beta (x) A(x),
\label{jacobian}
\end{equation}
with:
\begin{equation}\label{Aop}
A(x)=\frac{3g^2}{32\pi^2}G_{\mu\nu}^a\tilde G^{\mu\nu}_a=\frac{3g^2}{16\pi^2}\mbox{Tr}_c G_{\mu\nu}\tilde G^{\mu\nu},
\end{equation}
and where we have denoted $\beta(x)=\frac{1}{2}\sqrt{\frac{2}{3}}\alpha_A^0(x)$, $\tilde G^a_{\mu\nu}=\epsilon_{\mu\nu\alpha\beta}G^{\alpha\beta, a}$ and $G_{\mu\nu}=G^a_{\mu\nu}\lambda_a/2$. Recall that, since the strong coupling constant $g$ scales as $1/\sqrt{N_c}$, the anomaly term is proportional to $1/N_c$ so it vanishes as $N_c\rightarrow \infty$. In addition, the  factor of 3 in the numerator of~\eqref{Aop} comes from the trace over flavor space. Note that if we take (\ref{master}) and (\ref{ST}) with $a=0$ and  $\mathcal{O}=1$ (or any other operator invariant under $U_A(1)$ transformations), we recover the familiar anomalous equation for the abelian axial current (understood in the sense of expectation values): 
\begin{equation} \label{anomalyeq}
\partial_\mu J_5^\mu=2i\bar\psi\mathcal{M}\gamma_5\psi+A(x),
\end{equation}
with $J_5^\mu=\bar\psi \gamma_5\gamma^\mu\psi$.

Therefore, integrating \eqref{master} over the whole space-time and changing to the Euclidean, we have:

\begin{equation}
\left\langle\bar\psi(y)\left\{\frac{ \lambda^a}{2},\frac{ \lambda^b}{2}\right\}\psi(y)\right\rangle=-\frac{1}{2}\int_T dx \left\langle P^b(y)\left[i\bar\psi(x)\{\frac{\lambda^a}{2},\mathcal{M}\}\gamma_5\psi(x)\right]\right\rangle \qquad (a=1,\dots 8, b=0,\dots, 8) 
\label{PS}
\end{equation} 

\begin{equation}
\left\langle\bar\psi(y)\lambda^b\psi(y)\right\rangle=-\int_T dx \left\langle P^b(y)\left[i\bar\psi(x)\mathcal{M}\gamma_5\psi(x)\right]\right\rangle - \frac{1}{2}\int_T dx \mean{P^b(y) A(x)} \qquad (b=0,8)
\label{PSanom}
\end{equation}

For all flavor indices $a$ and $b$, the left-hand side of the above equations will become a combination of quark condensates, 
while the right-hand side will turn into a combination of zero-momentum euclidean pseudoscalar correlators, i.e., a pseudoscalar susceptibilities, which we define as:
\begin{equation}
\chi_P^{ab}\equiv\int_T dx \left\langle P^a (x) P^b (y) \right\rangle=K_P^{ab}(p=0),
\label{pseudodef}
\end{equation} 
with $K_P^{ab}(p)$ the Fourier transform of the Euclidean pseudoscalar correlator $K_P^{ab}(x-y)=\left\langle P^a (x) P^b (y)\right\rangle$, and so on for the anomalous susceptibilities:

\begin{align}
\chi_P^{bA}&\equiv\int_T dx \left\langle P^b (x) A (y) \right\rangle \qquad b=0,8,\\
\chi_P^{AA}&\equiv\int_T dx \left\langle A (x) A (y) \right\rangle.
\end{align}

Once we have established the main setup, we will discuss in detail the different cases of interest, which correspond to the different physical channels, starting from the simplest two-flavor case. 
In turn, we will verify the  obtained identities within the low-energy representation of QCD provided by $SU(3)$ and $U(3)$ ChPT, including finite temperature $T$ corrections at the order considered.  
All the equations above are formulated so that the extension to finite $T$ can be performed through the corresponding change in the correlation functions. 
Actually, to the order we are considering here, all temperature corrections will show up through the $T$-correction to the tadpole function coming from the finite part of the meson propagators at equal space-time points (we follow the same dimensional regularization scheme  as in \cite{Gasser:1984gg}):

\begin{align}
\mu_i(T)=\frac{M_{0i}^2}{32\pi^2 F^2}\log\frac{M_{0i}^2}{\mu^2}+\frac{g_1(M_{0i},T)}{2F^2}, \\
g_1(M_{0i},T)=\frac{T^2}{2\pi^2}\int_{M_{0i}/T}^\infty dx  \frac{\sqrt{x^2-(M_{0i}/T)^2}}{e^{x}-1},
\label{mudef}
\end{align}
with $M_{0i}$ the tree-level mass of the meson, $F$ the pion decay constant in the chiral limit and $\mu$  the renormalization scale. For the tadpole thermal functions $g_1$, which vanish at $T=0$,  we follow the same notation as in \cite{Gerber:1988tt}.

\subsection{Two-flavor case}
Here, we will reobtain the identity already analyzed in \cite{Nicola:2013vma}.  In SU(2), the generators of the algebra $\lambda^a$ are the Pauli matrices, 
so that in the isospin limit $m_u=m_d=\hat m$, $\{\frac{1}{2} \lambda^a,\mathcal{M}\}=\hat m \lambda^a$ and $\left\{ \lambda^a, \lambda^b\right\}=2\delta^{ab}$ and ~\eqref{PS} reduces to:
\begin{equation}
\delta^{ab}\condl=-\hat m \chi_P^{ab},
\label{PSSU2}
\end{equation}
with $\condl=\left\langle \bar uu +\bar dd \right\rangle$ the light quark condensate. 

To verify the identity \eqref{PSSU2} in $SU(2)$ ChPT, we calculate next the $\chi_P$ by coupling external pseudoscalar sources $p^a$ to the ChPT Lagrangian~\cite{Gasser:1983yg} and differentiating it with respect to them. 
Including finite-temperature corrections we have to NLO (one-loop)~\cite{Nicola:2013vma}:
\begin{align}
\chi_P^{ab} (T)&=\delta^{ab}\frac{4B_0^2F^2}{M_{0\pi}^2}\left[1+\frac{2M_{0\pi }^2}{F^2}(l^r_3+h_1^r)-3\mu_\pi(T)\right]+\Od(F^{-2})=-\delta^{ab}\frac{\condl(T)}{\hat m}+\Od(F^{-2}),
\end{align}
where $M_{0\pi}^2=2B_0\hat m$, $l_3^r$ and $h_1^r$ are renormalized scale-dependent low-energy constants (LECs)~\cite{Gasser:1983yg}  and the finite-$T$ quark condensate was derived at this order first in \cite{Gasser:1986vb}. The $\Od(F^{-2})$ encodes the NNLO corrections. The scale dependence of the LECs above is such that  $\chi_P^{ab} (T)$ is scale independent. Therefore, the identity \eqref{PSSU2} holds in $SU(2)$ ChPT up to NLO. 

\subsection{Three-flavor case}

Let us analyze separately the channels corresponding to the quantum numbers of the $\pi$, $K$ and $\eta-\eta'$, corresponding to all possible values of $a,b$ in eqs. \eqref{PS} and \eqref{PSanom}. As explained above,  anomalous contributions will enter naturally in the $\eta-\eta'$ sector. 

\subsubsection{$\pi$-channel ($a,b=1,2,3$)} 

In this case, $\{\frac{1}{2} \lambda^a,\mathcal{M}\}=\hat m \lambda^a$ and $\displaystyle \left\{\frac{ \lambda^a}{2},\frac{ \lambda^b}{2}\right\}=\delta^{ab}\left[\frac{\ID}{3}+\frac{1}{\sqrt{3}}\frac{ \lambda^8}{2}\right]$, 
so~\eqref{PS} reads:
\begin{equation}
\delta^{ab} \condl=-\hat m  \chi_P^{ab} \qquad (a=1,2,3),
\label{PS123}
\end{equation}
%\begin{equation}\label{PS123}
%\delta^{ab}\left\langle\bar uu(x)+\bar dd(x)\right\rangle=-\hat m \mean{P^b(y)P^a(x)},
%\end{equation}
which is precisely the relation obtained for the two-flavor case in \eqref{PSSU2}. 
Even for three flavors, the light sector identity decouples,  i.e, it does not include the strange condensate.  
Note that in the degenerate $SU(3)$ limit, i.e. for $m_u=m_d=m_s=m_q$ (we distinguish it from $\hat m$, the light mass in the non-degenerate case),
the quark condensates degenerate $\condu=\condd=\conds$ and the previous expression reduces to:
\begin{equation}\label{PSdeg123}
\frac{2}{3}\delta^{ab} \mean{\bar\psi\psi} =-m_q \chi_P^{ab}  \qquad (a=1,2,3 \quad \mbox{degenerate limit}),
\end{equation}
with $\mean{\bar\psi\psi}=\left\langle\bar uu+\bar dd+\bar ss\right\rangle$.
As we will discuss below, the study of the degenerate limit is of interest since it will allow us to test that the correlators corresponding to the octet members obey the same transformation rule, 
while the singlet transforms differently.

Now, we proceed to the verification of \eqref{PSdeg123} with the representation provided by $SU(3)$ ChPT, which, for the Ward identities involving the octet pseudoscalar correlators, is the most general low-energy framework  involving pions, kaons and the octet $\eta_8$. Nevertheless, as we will see below, this formalism will have to be extended to $U(3)$ when evaluating the singlet operator.
Hence, we consider the $SU(3)$ chiral Lagrangian up to fourth order in derivatives with an external pseudoscalar source \cite{Gasser:1984gg}. 
Similarly to the $SU(2)$ calculation in \cite{Nicola:2013vma}, we derive the pseudoscalar susceptibility up to NLO (one loop) and prove that the relation~\eqref{PS123} is also satisfied, namely:
\begin{align}\label{chptI3}
\chi_P^{ab}(T)=&\delta^{ab}\frac{4B_0^2F^2}{M_{0\pi}^2}\left\{1+\frac{4}{F^2}\left[\left(H_2^r + 4 L_6^r + 2 L_8^r\right) M_{0\pi}^2+8 L_6^r M_{0K}^2  \right]-3\mu_\pi(T) -2\mu_K(T)-\frac{1}{3}\mu_\eta(T) \right\}+\Od(F^{-2}) \nonumber\\ 
=&-\delta^{ab}\frac{\condl(T)}{\hat m}+\Od(F^{-2}) \qquad (a,b=1,2,3),
\end{align}
with $M_{0K}^2=B_0(m+m_s)$ and  $L_6^r$, $L_8^r$ and $H_2^r$ the renormalized $SU(3)$ LECs~\cite{Gasser:1984gg}, 
such that the condensates and susceptibilities above are scale independent.  
The explicit expressions for the  $SU(3)$ quark condensates at this order can be found for instance in \cite{GomezNicola:2010tb} and \cite{GomezNicola:2012uc} for the $T=0$ and $T\neq0$ case, respectively.

\subsubsection{$K$ channel ($a,b=4,5,6,7$)} 

For $a=4,5,6,7$, we have now $\{\frac{1}{2} \lambda^a,\mathcal{M}\}=\frac{\hat m+m_s}{2} \lambda^a$ . Furthermore: 
\begin{align}
\left\{\frac{ \lambda^a}{2},\frac{ \lambda^b}{2}\right\}=\delta^{ab}\left\{\left[\frac{\mathbb{1}}{3}-\frac{1}{2\sqrt{3}}\frac{ \lambda^8}{2}\right]+\frac{1}{2}\frac{\lambda^3}{2}\right\} \qquad a,b=4,5,\\
\left\{\frac{ \lambda^a}{2},\frac{ \lambda^b}{2}\right\}=\delta^{ab}\left\{\left[\frac{\mathbb{1}}{3}-\frac{1}{2\sqrt{3}}\frac{ \lambda^8}{2}\right]-\frac{1}{2}\frac{\lambda^3}{2}\right\} \qquad a,b=6,7.
\end{align}

Note that the terms above proportional to $\lambda^3$ will not contribute in the isospin limit that we are considering here, since $\condu=\condd$. 
Therefore in this case the relation~\eqref{PS} takes the form:
\begin{equation}\label{PS4567}
\delta^{ab}\left[\condl+2\conds\right] =-\left(\hat m+m_s\right) \chi_P^{ab} \qquad (a,b=4,5,6,7),
\end{equation}
which in the degenerate case reduces again to (\ref{PSdeg123}) as we obtained in the $\pi$ channel: 
\begin{equation}\label{PSdeg4567}
\delta^{ab}\frac{2}{3}\mean{\bar\psi\psi}=-m_q \chi_P^{ab} \qquad (a,b=4,5,6,7 \quad \mbox{degenerate limit}).
 \end{equation}

Once again, we proceed to the calculation of the pseudoscalar susceptibility in this channel in $SU(3)$ ChPT at NLO and we obtain that the identity~\eqref{PS4567} holds as well, namely:
\begin{align}\label{chptI4}
\chi_P^{ab}(T)=&\delta^{ab}\frac{4B_0^2F^2}{M_{0K}^2}\left\{1+\frac{4}{F^2}\left[\left(H_2^r + 8 L_6^r + 2 L_8^r\right) M_{0K}^2+4 L_6^r M_{0\pi}  \right]-\frac{3}{2}\mu_\pi (T) -3\mu_K (T)-\frac{5}{6}\mu_\eta (T)\right\} \\
=&-\delta^{ab}\frac{\condl (T)+2\conds(T)}{\hat m+m_s}\qquad (a,b=4,5,6,7),
\end{align}
where, as in the previous cases, the scale dependence of the LECs is canceled with that in the $\mu_i$ contributions to render scale-independent results.

\subsubsection{$\eta-\eta'$ sector ($a=b=0,8$ and anomaly terms)} 

The physical mixing of the octet $\eta_8$ and singlet $\eta_0$ states will show up also in the  Ward identities for the corresponding pseudoscalar operators. Therefore, on the one hand, the low-energy representation requires  to introduce the singlet field consistently with the low-energy counting, at the same footing as the other pseudo-Goldstone fields. On the other hand, the identities involving the singlet ($a=0$)  leads to the  presence of the $U_A(1)$ anomalous jacobian $\mathcal{J}$, as given by \eqref{jacobian} and \eqref{PSanom}. 

Let us then consider first the case $a=b=8$, for which we have:
\begin{align}\label{M8}
\left\{\frac{\lambda^8}{2} ,\mathcal{M}\right\}=&\frac{1}{\sqrt{3}}\diag{\left(\hat m,\hat m,-2m_s\right)}=\frac{\hat m+2m_s}{3} \lambda^8+\frac{\sqrt 2}{3}(\hat m-m_s)\lambda^0,\nonumber\\
\left\{\frac{\lambda^8}{2},\frac{\lambda^8}{2}\right\}=&\left[\frac{\mathbb{1}}{3}-\frac{\sqrt{3}}{3}\frac{ \lambda8}{2}\right].
\end{align}
Thus, in this case the relation \eqref{PS} leads to:
\begin{equation}\label{PS88}
\condl+4\conds=
%\left\langle\bar uu+\bar dd+4\bar ss \right\rangle=
-\left(\hat m+2m_s\right)\chi_P^{88}-\sqrt 2(\hat m-m_s)\chi_P^{80},
\end{equation}
so, as already anticipated, the last term mixes the octet and singlet pseudo-scalar currents. 

Recall that the mixing term above is proportional to the difference $\hat m-m_s$, and then in the degenerate case we recover the same relation as for the previous isospin channels:
\begin{equation}\label{PSdeg88}
\frac{2}{3}\mean{\bar\psi\psi}=-m_q\chi_P^{88} \qquad \mbox{(degenerate limit)},
 \end{equation}
consistently with the idea that in the SU(3) limit, all members of the octet transform in the same way and the $\eta_8-\eta_0$ mixing angle vanishes in that limit ~\cite{DiVecchia:1980ve,Guo:2011pa,Guo:2012ym,Guo:2012yt,Guo:2015xva}.

Before discussing the low-energy representation of~\eqref{PSdeg88}, let us now derive the rest of relations in this sector. 
Let us take now  $a=8, b=0$, so the transformation (\ref{fermiontrans}) is still not anomalous. Since 
$$\left\{\frac{ \lambda^8}{2},\frac{ \lambda^0}{2}\right\}=\frac{1}{2}\sqrt{\frac{2}{3}}\lambda^8,$$
the relation~\eqref{PS} reads now:
\begin{equation}\label{PS80}
\condl-2\conds=
%\mean{\bar uu+\bar dd-2\bar ss}=
-\frac{(\hat m+2m_s)}{\sqrt{2}}\chi_P^{80}-(\hat m-m_s)\chi_P^{00},
\end{equation}
which in addition to the octet-singlet mixing involves the singlet-singlet correlator. 
Furthermore, in the degenerate case, \eqref{PS80} implies actually the vanishing of the singlet-octet susceptibility: 
\begin{equation}\label{PSdeg80}
\chi_P^{80}=0 \qquad \mbox{(degenerate limit)}.
\end{equation}

We will end this analysis considering the cases where  $a=0$ and thus, where the jacobian is not zero. We have to use then~\eqref{PSanom}, which in particular for $b=8$  and using 
\begin{equation}
\mathcal{M}=\frac{1}{\sqrt{3}}(m-m_s)\lambda^8+\frac{1}{\sqrt{6}}(2m+m_s)\lambda^0,
\end{equation}
 leads to:

\begin{equation}\label{PS08}
\condl-2\conds=
%\mean{\bar uu+\bar dd-2\bar ss}=
-(\hat m-m_s)\chi_P^{88}-\frac{(2\hat m+m_s)}{\sqrt 2}\chi_P^{80}-\frac{\sqrt{3}}{2} \chi_P^{8A}.
\end{equation}

Note that the matrix element $ \left\langle P^0 A\right\rangle$ is nonzero due to the mixing of the anomaly and the singlet, which have the same quantum numbers. Nevertheless, it vanishes in the $N_c\rightarrow\infty$ limit, 
where the anomaly is absent. In addition, $ \left\langle P^8A\right\rangle$  is different from zero  due to the mixing  between the octet and singlet currents. 
However in the degenerate case, the octet-singlet mixing angle is zero and thus,  in the exact $SU(3)$ limit or in $N_c\rightarrow\infty$ limit, the anomaly term in \eqref{PS08} vanishes.
Actually, in the degenerate case, using \eqref{PSdeg80}, we get from \eqref{PS08}:

%\begin{equation}
%\partial_\mu\left\langle\mathcal{O}(x_1,\cdots,x_n)J_A^{0\mu}(x)\right\rangle=\frac{g^2}{16\pi^2}\left\langle \mathcal{O}(x_1,\cdots,x_n) Tr\left[G_{\mu\nu}\tilde G^{\mu\nu}\right]\right\rangle,
%\end{equation}

% \begin{equation}
%\partial_\mu\left\langle P^8(y)J_A^{0\mu}(x)\right\rangle=\frac{g^2}{16\pi^2}\left\langle P^8(y) Tr\left[G_{\mu\nu}\tilde G^{\mu\nu}\right]\right\rangle=\mathcal M_0\mean{P^8(x)\mathcal O_A(y)},
%\end{equation}
%where $\mathcal M_0$ is a constant which scales as $1/N_c$. 
\begin{equation}\label{PSdeg8A}
\chi_P^{8A}=0 \qquad \mbox{(degenerate limit)},
\end{equation}
consistently with our previous comments. 
Note also that  the l.h.s of eqs.~\eqref{PS80}~and~\eqref{PS08} are exactly the same, which implies: 
\begin{equation}
\label{PSnonet}
(\hat m-m_s)\left(\chi_P^{00}-\chi_P^{88}\right)=\frac{\parent{\hat m-m_s}}{\sqrt 2}\chi_P^{80}+\frac{\sqrt{3}}{2}\chi_P^{8A},
\end{equation}
which for $N_c\rightarrow\infty$ implies: 
$$\chi_P^{80}=\sqrt{2}\left(\chi_P^{00}-\chi_P^{88}\right)\qquad (N_c\rightarrow\infty).$$ 

%\begin{equation}
%(m-m_s)\parent{\mean{P^0(y)P^0(x)}-\mean{P^8(y)P^8(x)}}=\frac{\parent{m-m_s}}{\sqrt 2}\mean{P^8(y)P^0(x)}+3\sqrt 2\mathcal M_0\mean{P^8(x)\mathcal O_A(y)},
%\end{equation}

Finally, considering \eqref{PSanom} for $b=0$, following the same steps as above and taking into account that $\left\{\frac{ \lambda^0}{2},\frac{ \lambda^0}{2}\right\}=\frac{1}{3}\mathbb{1}$,
we obtain:
\begin{equation}\label{PS00}
\condl+\conds=
%\mean{\bar uu+\bar dd+\bar ss}=
-\frac{\left(2\hat m+m_s\right)}{2}\chi_P^{00}-\frac{\left(\hat m-m_s\right)}{\sqrt{2}}\chi_P^{80}-\frac{\sqrt{6}}{4}\chi_P^{0A},
\end{equation}
%\begin{equation}\label{PS00}
%\mean{\bar uu(x)+\bar dd(x)+\bar ss(x)}=-\frac{\left(2m+m_s\right)}{2}\mean{P^0(y)P^0(x)}-\frac{\left(m-m_s\right)}{\sqrt{2}}\mean{P^0(y)P^8(x)}+3\mathcal M_0\mean{P^0(x)\mathcal O_A(y)},
%\end{equation}
which in the degenerate case reduces to:
\begin{equation}
\label{PS00deg}
\frac{2}{3}\mean{\bar\psi\psi}=-m_q\chi_P^{00}-\frac{1}{\sqrt{6}}\chi_P^{0A}  \qquad \mbox{(degenerate limit)}.
\end{equation}
Note that the result above  differs from the previous octet isospin channels in \eqref{PSdeg123}, \eqref{PSdeg4567}, \eqref{PSdeg88}, 
pointing out that the singlet transforms differently than the octet due to its mixing with the $U_A(1)$ anomaly. Actually, for $N_c\rightarrow\infty$ we obtain consistently that all members of the nonet transform in the same way. 

So far we have obtained four equations for this sector, namely~\eqref{PS88}, \eqref{PS80}, \eqref{PS08} and \eqref{PS00}, in term of five pseudo-scalar susceptibilities: 
$\chi_P^{88}$,  $\chi_P^{00}$, $\chi_P^{80}$, $\chi_P^{0A}$ and $\chi_P^{8A}$. It is easy to check that the rank of this system of equations is 4, which allows to express four of the $\chi_P$ in terms of only one and combinations of quark condensates. 

For completeness, we will also include in the system of equations a relations for $\chi_P^{AA}$. This can be done by considering eqs.~(\ref{master}) and (\ref{ST}) with $\mathcal{O}(y)=A(y)$ and $a=0$ and 8, 
which gives two additional equations. Since that operator is invariant under any fermion transformation, the first term in (\ref{master}) vanishes and using our previous results we obtain:
\begin{align}
\mathcal{O}=A, a=8 \rightarrow  \chi_P^{8A}&=\sqrt{2}\frac{m_s-\hat m}{(\hat m+2m_s)}\chi_P^{0A},\label{PS8A}\\
\mathcal{O}=A, a=0 \rightarrow \chi_P^{AA}&=-\frac{2}{\sqrt{3}}(\hat m-m_s)\chi_P^{8A}-\frac{2}{\sqrt{6}}(m_s+2\hat m)\chi_P^{0A}.
\label{PS0A}
\end{align}
In the degenerate limit, (\ref{PS8A}) gives $\chi_P^{8A}=0$, consistently with (\ref{PSdeg8A}),  while (\ref{PS0A}) reduces to:
\begin{equation}
\chi_P^{AA}=   -\sqrt{6}m_q\chi_P^{0A} \qquad \mbox{(degenerate limit)}.
\label{PSdegAA}
\end{equation}

Combining now~\eqref{PS88}, \eqref{PS80}, \eqref{PS08} and \eqref{PS00} with \eqref{PS8A} and \eqref{PS0A}, we obtain a system of six equations with rank 5 and six unknowns, 
whose solution can be written as:
\begin{align}
\chi_P^{88}=&-\frac{1}{3}\left(\frac{\condl}{\hat m}+\frac{4\mean{\bar ss}}{m_s}\right)+\frac{\sqrt{3}}{9}\frac{\hat m-m_s}{\hat m m_s}\chi_P^{8A},\nonumber\\
\chi_P^{80}=&-\frac{\sqrt 2}{3}\left(\frac{\condl}{\hat m}-\frac{2\mean{\bar ss}}{m_s}\right)-\frac{\sqrt{6}}{18} \frac{\hat m+2m_s}{\hat m m_s} \chi_P^{8A},\nonumber \\
\chi_P^{00}=&-\frac{2}{3}\left(\frac{\condl}{\hat m}+\frac{\mean{\bar ss}}{m_s}\right)-\frac{\sqrt{3}}{18}\frac{(\hat m+2m_s)^2}{\hat m m_s (m_s-\hat m)} \chi_P^{8A},\nonumber\\
\chi_P^{0A}=&\frac{\sqrt{2}}{2}\frac{\hat m + 2m_s}{m_s-\hat m} \chi^{8A},\nonumber\\
\chi_P^{AA}=&-3\sqrt{3}\frac{\hat m m_s}{m_s-\hat m} \chi^{8A}.
\label{finaleqseta}
\end{align}

In case we would have not included~\eqref{PS8A}~and~\eqref{PS0A}, we would have obtained the same set of equations~\eqref{finaleqseta} but without the last one. 
These equations, together with those previously obtained in the pion and kaon sectors, namely \eqref{PS123} and \eqref{PS4567}, constitute some of the more important results of this work. 
We recall that in previous analyses \cite{Broadhurst:1974ng,Bochicchio:1985xa} only results for the octet in the SU(3) limit were discussed. 

%Recall that dimensionally in energy, the susceptibilities $\chi_P^{ab}$ have dimension 2, while $\chi_P^{aA}$ have dimension 3 and $\chi_P^{AA}$ has dimension 4. The equations are dimensionally %correct because the quark condensates have dimension 3 and the quark masses dimension 1. 

We remark once more that  we have obtained these Ward identities formally  within QCD, but they have to be verified with explicit representations of the operators involved, which we are carrying out here using the low-energy representation provided by ChPT. Therefore, it remains to verify equations \eqref{finaleqseta} in that framework. 
For that purpose, we have to include consistently the $\eta^\prime$ field, which saturates the singlet current. This can be done by considering the $U(3)$ extension of ChPT in the large $N_c$ limit~\cite{'tHooft:1973jz,Witten:1979kh,Rosenzweig:1979ay,Witten:1979vv,Coleman:1980mx,Veneziano:1980xs,DiVecchia:1980ve,Witten:1980sp,HerreraSiklody:1996pm,Kaiser:2000gs}, 
since the  $U_A(1)$ anomaly vanishes and the singlet field $\eta_0$ becomes the ninth pseudo-Goldstone boson for $N_c\rightarrow\infty$. Its mixing with the octet $\eta_8$ yields the physical $\eta$-$\eta'$ fields. The standard chiral counting in meson masses, energies and temperatures is extended then to include the $1/N_c$ counting, so that generically the expansion is performed in a parameter $\delta$ such  that $M_k^2, E_k^2, T^2,\hat m,m_s =\Od(\delta)$ and $1/N_c=\Od(\delta)$. In this counting,  $F^2=\Od(N_c)=\Od(1/\delta)$, which suppresses loops, while the counting of the different  LECs, 
according to their $\Od(N_c)$  trace structure, is given in detail in \cite{HerreraSiklody:1996pm,Guo:2015xva}. 

We have calculated all the pseudoscalar susceptibilities involved in this sector, as well as the quark condensates and we have verified the identities \eqref{finaleqseta} up to NNLO in the $\delta$ expansion.  The LO is $\Od(\delta^{-2})$ for $\chi_P^{88,80,00}$, $\Od(\delta^{-1})$ for $\chi_P^{0A,8A}$ and $\mean{\bar q_i q_i}$ and $\Od(1)$ for $\chi^{AA}$. Apart from including pseudoscalar sources $p^a$ in the $U(3)$ effective Lagrangian, as indicated for instance in \cite{HerreraSiklody:1996pm,Guo:2015xva}, we have also included the anomalous external field $\theta(x)$, which couples to the  QCD Lagrangian through the term

\begin{equation}
\label{thetalag}
{\cal L}_\theta=-\frac{1}{6}\theta(x)A(x),
\end{equation}
so that the anomalous change produced by the jacobian \eqref{jacobian} is compensated by a change $\theta(x)\rightarrow \theta(x)-6\beta(x)$ \cite{Gasser:1984gg,HerreraSiklody:1996pm} in the QCD effective lagrangian. In this way, the expectation values involving the anomaly, such as those appearing in $\chi_P^{aA}$, $\chi_P^{AA}$, can be derived as $\mean{A\cdots}=-6\frac{\delta}{\delta \theta(x)}\cdots\log Z$ with $Z$ the Euclidean generating functional. In the effective lagrangian, $\theta(x)$ couples through the operator $X=\log\det U + i\theta(x)$ with  $U$ the NGB matrix field \cite{HerreraSiklody:1996pm}.

The order we are calculating here requires to consider the effective Lagrangians up to NNLO, namely ${\cal L}_\delta^0, {\cal L}_\delta, {\cal L}_{\delta^2}$ in the notation of \cite{Guo:2015xva}, as well as the NLO and NNLO corrections to the self-energies of all meson propagators, including the $\eta$ and $\eta'$ ones. Recall that those self-energy corrections for the octet fields $\pi,K,\eta_8$ differ from those calculated in $SU(3)$ ChPT due to the $\eta_0$ loops. Besides, the $\eta-\eta'$ mixing angle has to be incorporated at the relevant order. All these ingredients, Lagrangians, self-energies and the mixing angle, are given in detail in the recent work \cite{Guo:2015xva}.

These results for pseudoscalar susceptibilities and quark condensates are presented in this work for the first time in the $U(3)$ ChPT framework. Nevertheless, since they are rather long expressions, we collect them in Appendix \ref{sec:app}. The light and strange quark condensates are given in \eqref{lightchptU3} and \eqref{strangechptU3} and $\chi_P^{8A}$ in \eqref{chptchi8A}, while the rest of susceptibilities in this sector can be read directly from those results and equations \eqref{finaleqseta}. 
A very nontrivial check of consistency of our results, apart from them satisfying the Ward identities \eqref{finaleqseta},  is that they remain finite and scale-independent with the renormalization of the $U(3)$ LECs given for instance in~\cite{Kaiser:2000gs}. 
Recall that such LEC renormalization is genuinely different from the standard $SU(3)$ one in \cite{Gasser:1984gg} due to the appearance of new LECs, 
as well as the modification of the old ones from $\eta'$ loops, which in particular requires the renormalization of the $B_0$ constant given also in ~\cite{Kaiser:2000gs}. 

Finally, also for completeness, we have verified that the identities obtained before for the $\pi$ and $K$ sector, namely \eqref{PS123} and \eqref{PS4567},  also hold  within the $U(3)$ ChPT formalism up to NNLO in the $\delta$ expansion, calculating explicitly the modified $\chi_{P}^{ab}$ for $a,b=1,\dots,8$. Thus, the quark condensates in the l.h.s. of those equations are modified, as given by \eqref{lightchptU3} and \eqref{strangechptU3} in Appendix \ref{sec:app}, and the $\chi_{P}^{ab}$ in the r.h.s. change accordingly so that the Ward identities  hold.

\section{Lattice data: pseudoscalar Ward identities and scaling of screening masses}
\label{sec:latt}

In this section we will analyze lattice data which support the previous Ward identities that we have obtained in QCD and ChPT for $N_f=2 $ and $N_f=3$.  
First, we will comment on recent lattice results which  compare directly the pseudoscalar susceptibilities with the corresponding condensate combination. 
This comparison will provide us with an estimate of the typical lattice errors expected in those identities due to the finite-size effects. Second, we will propose an interpretation of the scaling behavior of  lattice screening masses for pion, kaon and $\bar s s$ channels based precisely on these identities, hence extending the pion-channel results presented in \cite{Nicola:2013vma}. 

Before going on, let us notice that the lattice results for screening masses that we will analyze here are presented for the $\pi^+$,  $K^+$ and $\bar s s$ channels. 
The first two correspond to the identities \eqref{PS123} and \eqref{PS4567} respectively (we are assuming isospin symmetry) but the third one is a linear combination of the susceptibilities appearing in \eqref{finaleqseta}. 
Namely:
\begin{equation}
\chi_P^{\bar s s}=\frac{1}{3}\chi_P^{88}+\frac{1}{6}\chi_P^{00}-\frac{2}{3\sqrt{2}}\chi_P^{08},
\end{equation} 
which using \eqref{finaleqseta} gives:
\begin{equation}
\chi_P^{\bar s s}=-\frac{\mean{\bar ss}}{m_s}+\frac{\hat m}{4\sqrt{3}  m_s\left(\hat m -m_s\right)}\chi_P^{8A}.
\label{PSsbars}
\end{equation}

Two important features of the above equation will be relevant for our following analysis. 
First, the light condensate contribution disappears in this combination, which, as we will see, will play an important role in the screening mass description for this channel. 
Second, unlike the other anomalous contributions in \eqref{finaleqseta}, the anomalous term in \eqref{PSsbars} is weighted by a ${\hat m}/m_s$ factor, 
which leads to a suppression of that term in the physical case  $\hat m\ll m_s$. 

A very recent lattice analysis for domain-wall fermions \cite{Buchoff:2013nra} compares directly these relations for the $\pi^+$~\eqref{PS123} and $\bar ss$ channel \eqref{PSsbars}. 
However, the determination of the Ward identity for the axial current \eqref{master} in the lattice receives a correction which accounts for finite-size effects, corresponding to quadratic divergences in condensates that we will discuss below. 
These corrections are written in the form of additional compensating axial currents, which in the  $\bar s s$ channel carry also the anomalous part ~\cite{Furman:1994ky,Blum:2000kn}. On the one hand, neglecting this correction, the deviations of the identity are around a 30-40 \% below $T_c$ in the light sector~\eqref{PS123} and typically less than 10\% for the $\bar s s$ channel \eqref{PSsbars}  (see Fig.1 in \cite{Buchoff:2013nra}).  On the other hand, the agreement is almost exact in these two channels when the compensating current is included. 
Thus, we can reinterpret this correction as an estimate of the lattice finite-size effects to the continuum relations, which we will keep in mind for our analysis of screening masses. 
Actually, for this analysis and in order to avoid quadratic divergences, it will be more meaningful to use subtracted condensates in the lattice instead of the naive continuum expressions. Note that the previously commented suppression of the  anomalous contribution in \eqref{PSsbars} is consistent with these lattice results, since otherwise the lattice deviations in the $\bar s s$ channel, which include the anomalous contribution as well as finite-size effects, should be much larger.  Finally, let us remind that the identity for the kaon channel \eqref{PS4567} has not  been checked in the lattice yet. 
This channel would be of interest since it is the only one mixing the light and strange condensates. 
Below, we will provide an indirect check of this channel identity through the study of the screening mass scaling.

Now, let us explore in more detail the implications of the above relations for the behavior of the light- and strange-channel screening masses in the lattice. 
Lattice screening masses are defined as the coefficient of the exponential falloff of a correlator at zero frequency and large spatial distances $K_P\sim \exp(-M^{sc}\vert z\vert)$, corresponding to taking the $p=0$ limit as $(\omega=0,\vec{p}\rightarrow \vec{0})$.
In particular, for the pseudoscalar correlators defined in \eqref{pseudodef}, the most recent screening mass results in the $\pi^+$, $K^+$ channels are given in~\cite{Cheng:2010fe}, 
and in~\cite{Maezawa:2013nxa} for the $\bar s s$ one.
A prominent feature of those masses, clearly observed in lattice data, is that they  grow  near the chiral transition. 
Furthermore, this growing behavior is more pronounced for the $\pi^+$ channel than for the other two and slightly more for the $K^+$ channel than for the $\bar s s $ one (see e.g. Fig.2c in \cite{Cheng:2010fe}). 
Here, we provide  a natural explanation for this behavior in terms of the identities obtained in the previous section. 
The main idea behind this is that the sudden drop of the light condensate $\condl$ near the transition would be correlated with the mass growth through $M^2\sim \left[K_P(p=0)\right]^{-1}=\chi_P^{-1}\sim \condl^{-1}$ from \eqref{PS123} and so on for the other channels, where as we will see, the corrections due to the strange condensate  explain also the observed behavior. 

In principle, one would expect the susceptibilities to scale as the inverse of the pole mass squared from a parametrization of the form $K_P^{-1}(\omega,\vec{p})\sim -\omega^2+A^2(T)\vert\vec{p}\vert^2+M^{pole}(T)^2$,
thoroughly used in lattice analysis \cite{Karsch:2003jg},  with $A(T)=M^{pole}(T)/M^{sc}(T)$, being $M^{pole}$ and $M^{sc}$ the pole and screening masses respectively. 
Thus, $\chi_P(T)\sim \left[M^{pole}\right]^{-2}$ and the pole mass is understood as the counterpart of the screening mass. 
The difference between screening and pole masses parametrized by $A(T)$ comes from the different spatial and temporal dependence of  self-energies  in the thermal bath. 
However, lattice analysis do not measure the pole masses, so we must rely on the reasonable assumption of a soft $T$ behavior  $A(T)\sim 1$ below $T_c$. This is supported for instance by finite-$T$ ChPT, where those differences show up at the two-loop level and remain small up to  temperatures close to the transition \cite{Schenk:1993ru}.  In addition, $T$-dependent residues can also enter.  Actually, in general we should write $\chi_P=N_\chi/\left[M^2+\Sigma_T(0,0)\right]$ with $N_\chi$  a $T=0$ normalization, $M$  the tree level mass of the correlator and $\Sigma_T(\omega,\vec{p})$  its $T$-dependent self-energy. Expanding around $p=0$, $\Sigma(\omega,\vec{p};T)=\Sigma_T(0,0)+\alpha(T)\omega^2-\beta(T)\vert\vec{p}\vert^2+\Od(p^4)$ yields the above lattice parametrization with $A^2(T)=\left[1+\beta(T)\right]/\left[1+\alpha(T)\right]$ and $\left[M^{pole}\right]^2(T)=\left[M^2+\Sigma_T(0,0)\right]/\left[1+\alpha(T)\right]$. Therefore, $N_\chi\chi_P^{-1}(T)=\left[1+\alpha(T)\right]A^2(T)\left[M^{sc}(T)\right]^2$. The assumption that  residues are soft  near chiral restoration and that the relevant scaling is governed by the mass contribution, has been  followed also in the case of the scalar susceptibility $\chi_S$ in \cite{Nicola:2013vma}. In that case, saturating $\chi_S$ by the dynamical $f_0(500)$ thermal mass leads to a successful $T$-behavior, developing a peak close to the lattice prediction for the transition temperature. 

Assuming then that both the residue $\alpha(T)$ and $A(T)$ are smooth functions of temperature leads to the following predictions for the scaling of the screening masses in the $\pi$ and $K$ channels, according to the Ward identities \eqref{PS123} and \eqref{PS4567}:
\begin{align}
\frac{M^{sc}_\pi (T)}{M^{sc}_\pi (0)} \sim \left[\frac{\chi_P^\pi(0)}{\chi_P^\pi(T)}\right]^{1/2}=\left[\frac{\condl(0)}{\condl (T)}\right]^{1/2}\label{scalingpi}\\
\frac{M^{sc}_K (T)}{M^{sc}_K (0)}\sim \left[\frac{\chi_P^K(0)}{\chi_P^K(T)}\right]^{1/2}=\left[\frac{\condl (0)+2\mean{\bar s s}(0)}{\condl (T)+2\mean{\bar s s}(T)}\right]^{1/2}.
\label{scalingK}
\end{align}

For the $\bar s s$ channel, we should include also the anomalous part proportional to $\chi_P^{8A}$ in \eqref{PSsbars}, which gives rise to a scaling relation in which quark masses are not canceled. However, as stated above, it is reasonable to neglect the anomalous part as far as critical scaling is concerned, since it is suppressed in the chiral limit and  so is observed in lattice data.   Thus, we arrive to a simplified version for the scaling in that channel:

\begin{align}
\frac{M^{sc}_{\bar s s} (T)}{M^{sc}_{\bar s s} (0)}\sim \left[\frac{\chi_P^{\bar s s} (0)}{\chi_P^{\bar s s} (T)}\right]^{1/2}\sim \left[\frac{\mean{\bar s s}(0)}{\mean{\bar s s} (T)}\right]^{1/2}.
\label{scalings}
\end{align}

Our next step will be to test  the above scaling laws with lattice data, within the uncertainties already commented, 
i.e., due to lattice finite-size effects and our ignorance about the $T$-dependence of the $A(T)$ and $\alpha(T)$ functions. 
Some qualitative interesting conclusions can already be extracted just by looking at the behavior near the transition of the different condensates involved in the above relations. 
The light condensate would vanish at the transition (in the chiral limit) and thus, we expect a large growing behavior for $M^{sc}_\pi$ from \eqref{scalingpi}, as it is seen in lattice data. 
However, the presence of the strange condensate contribution in the kaon channel \eqref{scalingK} would prevent it from diverging. Therefore, a softer behavior than in the pion case is expected, 
as it is also observed in the lattice. Finally, the cancellation of the $\condl$ contribution in the $\bar s s$ channel in \eqref{scalings} explains also why the growth is even slower in that channel. 

\begin{figure}
\centerline{\includegraphics[width=14cm]{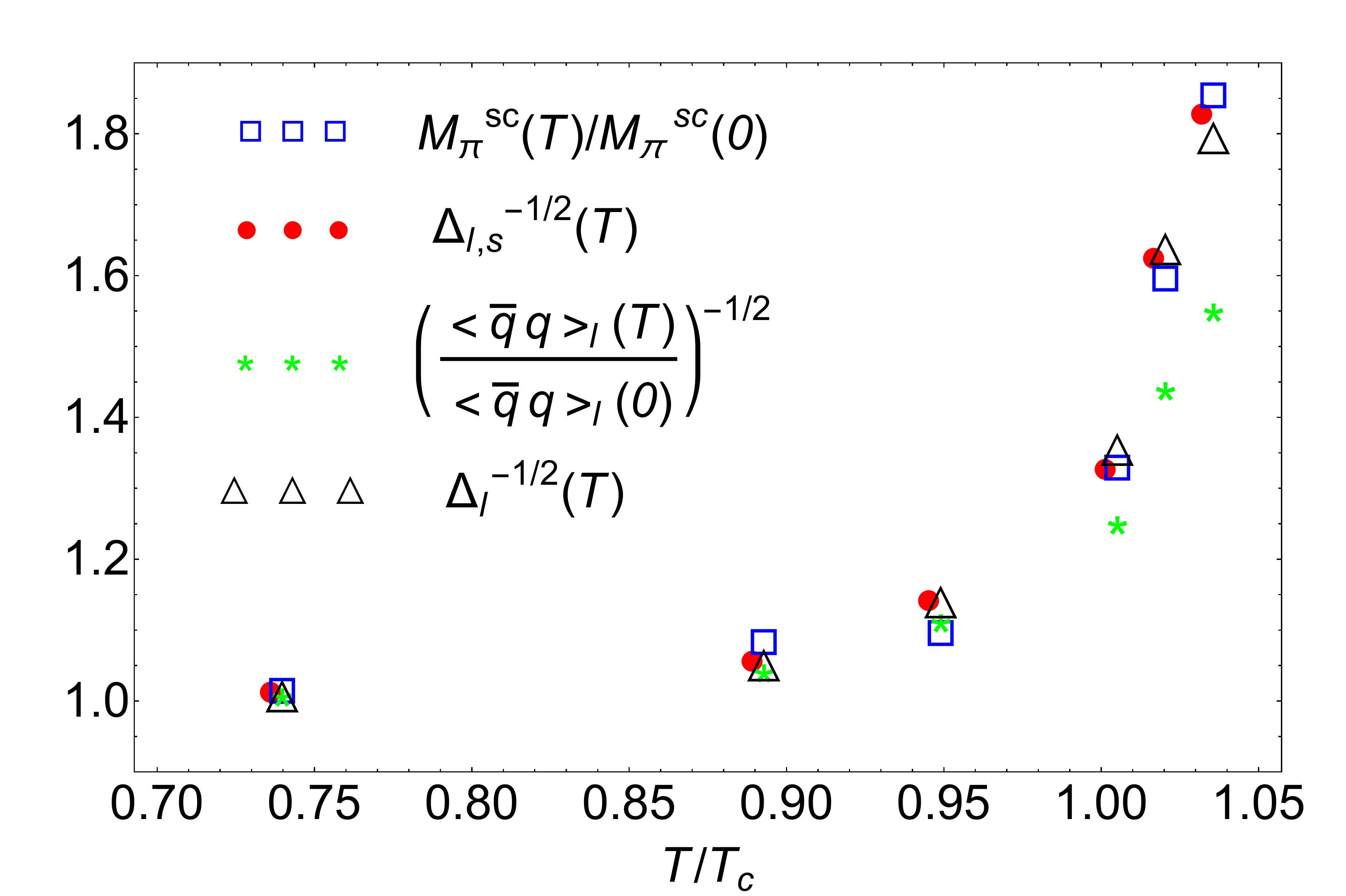}}
\caption{Comparison between the pseudoscalar screening mass ratio in the pion channel and the ratio of light condensates, subtracted and unsubtracted, defined in the main text. The  lattice data are taken from 
\cite{Cheng:2010fe} (masses) and \cite{Cheng:2007jq} (condensates) with the same lattice action and resolution and $T_c\simeq 196$ MeV. The values for $\Delta_l (T)$ correspond to the definition \eqref{Deltal} with $r_1^3\condl^{ref}=$ 0.749 and the value of $r_1\simeq$ 0.31 fm used in \cite{Cheng:2007jq}.}
\label{fig:corrpion}
\end{figure}

Let us  proceed now to a more quantitative analysis, considering first the pion channel, previously discussed in \cite{Nicola:2013vma}. 
In Fig.\ref{fig:corrpion}, we show the comparison of the screening masses for this channel taken from \cite{Cheng:2010fe} (blue squares) and the subtracted light condensate ratio $\Delta_{l,s}^{-1/2}$ taken from \cite{Cheng:2007jq} (red dots) and defined as:
\begin{equation}
\Delta_{l,s}(T)=\frac{\condl(T)-2\frac{\hat m}{m_s}\conds (T)}{\condl(0)-2\frac{\hat m}{m_s}\conds (0)}.
\label{Deltals}
\end{equation}
Both quantities are computed with the same lattice conditions, i.e. a $p4$ action with $N_\tau=6$ and $m_s=10\,\hat m$. 
The reason why the strange condensate is subtracted in \eqref{Deltals} is the presence of a lattice divergence proportional to $m_i/a^2$, 
with $a$ the lattice spacing, in the condensate $\mean{\bar q_i q_i}$. Thus, $\Delta_{l,s}(T)$ behaves as an order parameter for the chiral symmetry breaking in the lattice \cite{Bazavov:2009zn} 
in the same way that the light condensate $\condl$ does in the continuum. 
Furthermore, as explained above, this is precisely the type of quadratic divergence that requires a compensating axial current in the lattice calculation of the Ward identities. 
The difference in the continuum between $\Delta_{l,s}$ and $\condl(T)/\condl(0)$ is about 15\%  near $T_c$, estimated from NLO ChPT~\cite{Nicola:2013vma}. 
Note also that the lattice data in \cite{Cheng:2007jq} are somewhat outdated, in particular they predict a rather high value for $T_c\simeq$ 196 MeV. 
However, as stated before, there have been no updated results for screening masses in this channel with upgraded lattice conditions, as for instance those in \cite{Aoki:2009sc,Borsanyi:2010bp,Bazavov:2011nk,Buchoff:2013nra,Bhattacharya:2014ara}. 
For this reason, we will plot in our figures the results as functions of $T/T_c$, so the effect we are trying to put forward is emphasized independently of the accuracy of the lattice data. In addition, we will only take condensate data for those temperature values for which there are data for screening masses. 

The comparison between these two results in Fig.\ref{fig:corrpion} shows a clear correlation between them, which supports the scaling law in~\eqref{scalingpi}. 
More precisely, for those points showed in Fig.~\ref{fig:corrpion}, the maximum relative difference between $M^{sc}_\pi(T)/M^{sc}_\pi(0)$ and $\Delta_{l,s}^{-1/2}$ is about 4.3\% (third point) and less than 3\% for the others, 
which is highly remarkable, given the expected size of uncertainties mentioned above.
Thus, it  provides a natural explanation for the growth of the pion screening masses in terms of the quark condensate, despite the various uncertainties involved.
In the same figure, we also plot the scaling of the light condensate without the subtraction (green stars). The correlation is  worse, as expected from the lattice divergences already discussed; 
the maximum difference being now about 17\% in the last point, and less than 10\% for the others. 
As a matter of fact, to get more insight about this scaling law and the importance of considering subtracted condensates for lattice data, 
we consider another subtracted lattice order parameter $\Delta_l (T)$ alternative to $\Delta_{l,s}$ in \eqref{Deltals}, which, following ~\cite{Bazavov:2011nk}, we define as:
\begin{equation}
\Delta_l (T)=\frac{\condl(T)-\condl(0)+\condl^{ref}}{\condl^{ref}},
\label{Deltal}
\end{equation}
where $\condl^{ref}$ is some $T=0$ reference value, which in \cite{Bazavov:2011nk} is taken as  the value for $\condl$ in the chiral limit obtained by the MILC collaboration \cite{Bazavov:2009bb}. The above combination is free from the lattice divergences previously commented and behaves as an order parameter similarly to $\Delta_{l,s}$. The quantity $\Delta_l^R$ defined in \cite{Bazavov:2011nk} corresponds to the numerator of \eqref{Deltal} normalized to ensure renormalization-group invariance. However, since we are considering older lattice results with very different lattice conditions~\cite{Cheng:2007jq}, we cannot  take the same value for  $\condl^{ref}$ in \cite{Bazavov:2011nk}. 
What we do instead is to treat $\condl^{ref}$ as a fit parameter, minimizing the square sum difference between the data for $M^{sc}_\pi (T)/M^{sc} (0)$ in \cite{Cheng:2010fe} and $\Delta_l(T)^{-1/2}$ with the condensate values of \cite{Cheng:2007jq}. We use for the condensates the dimensionless quantity $r_1^3\mean{\bar q q}$ where $r_1\simeq$ 0.31 fm defined in lattice analysis to set the physical scale \cite{Cheng:2007jq,Bazavov:2009bb,Bazavov:2011nk}. 
We show in Fig.\ref{fig:corrpion} the results for $\Delta_l^{-1/2}$ (black triangles) with $r_1^3\condl^{ref}=$ 0.749. 
That value is obtained by fitting the three channels with $\condl^{ref}$ and $\conds^{ref}$ as fit parameters (see below). The behavior is very similar to that of $\Delta_{l,s}^{-1/2}$ as expected, reaching a maximum of about 4\% for the relative differences with the screening masses. Putting this condensate values in physical units gives $\condl^{ref}\simeq$ (560 MeV)$^3$, which is high compared to typical $T=0$ phenomenological estimates  \cite{Bazavov:2009bb,Colangelo:2010et} but once again it is more meaningful to compare with  the values quoted in \cite{Cheng:2007jq} for the condensate, namely $\condl (T=0)\simeq$ (590 MeV)$^3$, which is actually larger than  $\condl^{ref}$  as it should if we think of $\condl^{ref}$ as a typical chiral limit value.

\begin{figure}
\centerline{\includegraphics[width=14cm]{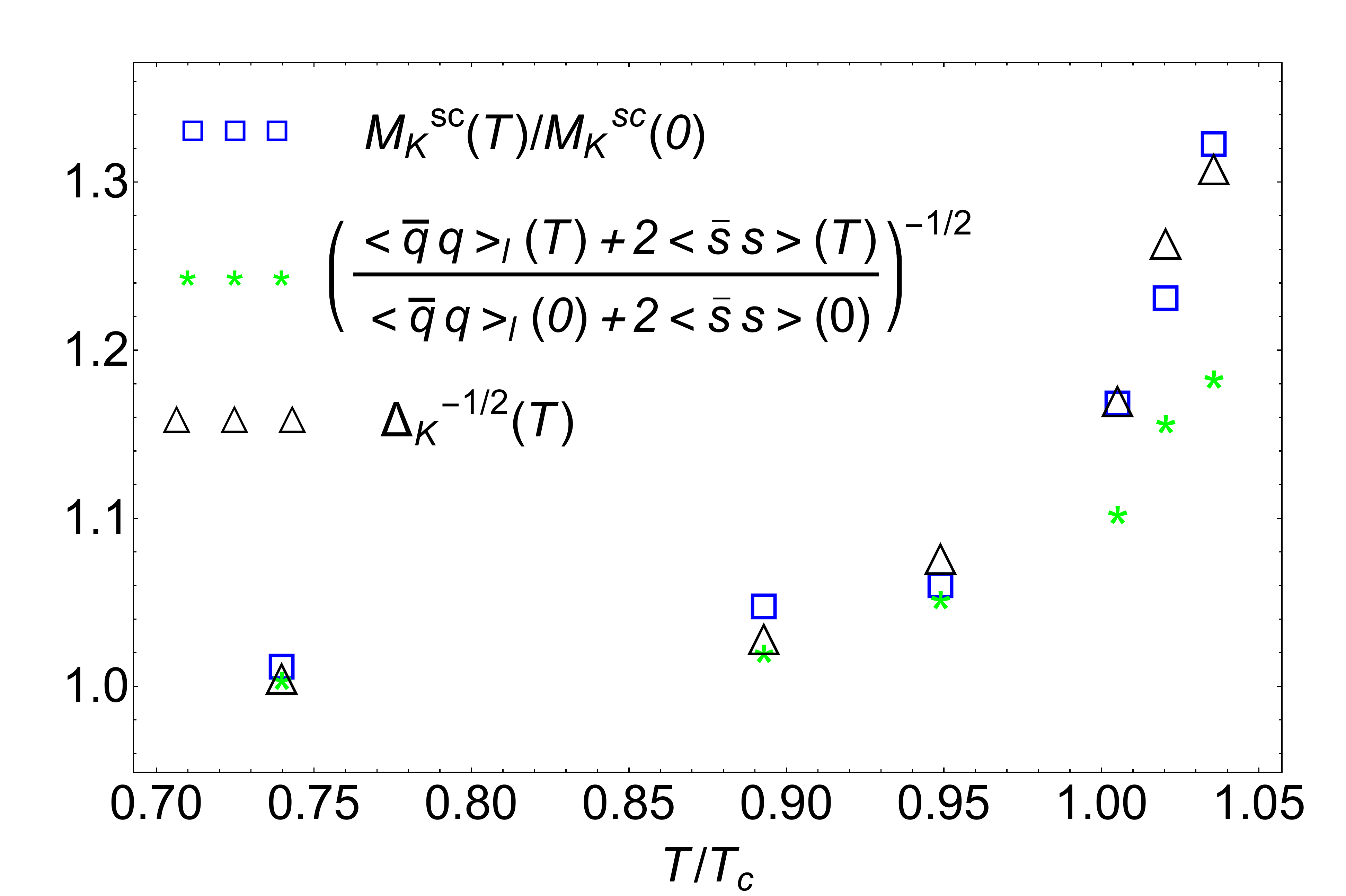}}
\caption{Comparison between the pseudoscalar screening mass ratio in the kaon channel and the ratios of light and strange condensate combinations, subtracted and unsubtracted, defined in the main text. The  lattice data are taken from 
\cite{Cheng:2010fe} (masses) and \cite{Cheng:2007jq} (condensates) with the same lattice action and resolution and $T_c\simeq 196$ MeV. The values for $\Delta_K(T)$ correspond to the definition \eqref{DeltaK} with $r_1^3\condl^{ref}=$ 0.749,  $r_1^3\conds^{ref}=1.109$ and the value of $r_1\simeq$ 0.31 fm used in \cite{Cheng:2007jq}.}
\label{fig:corrkaon}
\end{figure}
The study of the scaling law in the kaon and $\bar s s$ channels, \eqref{scalingK} and \eqref{scalings} respectively, can be worked out along similar lines. 
As it happened in the light channel, we expect that a simple comparison with the naive lattice condensates would give a worse correlation. 
Actually, lattice  divergences are proportional to $(\hat m+m_s)/a^2$ and $m_s/a^2$ in the kaon and $\bar s s$ channels respectively, hence enhanced by the strange mass. 
Once more, we consider subtracted condensates to eliminate those lattice divergences and to be able to study more accurately the proposed correlation. 
From our previous comments, we replace both $\condl$ and $\conds$ by their subtracted counterparts, so that we define, following the convention in \cite{Bazavov:2011nk}:
\begin{equation}
\Delta_K (T)=\frac{\condl(T)-\condl(0)+2\left[\conds(T)-\conds(0)\right]+\condl^{ref}+\conds^{ref}}{\condl^{ref}+\conds^{ref}},
\label{DeltaK}
\end{equation}

\begin{equation}
\Delta_s (T)=\frac{2\left[\conds(T)-\conds(0)\right]+\conds^{ref}}{\conds^{ref}}.
\label{Deltas}
\end{equation}

\begin{center}
\begin{figure}
\centerline{\includegraphics[width=14cm]{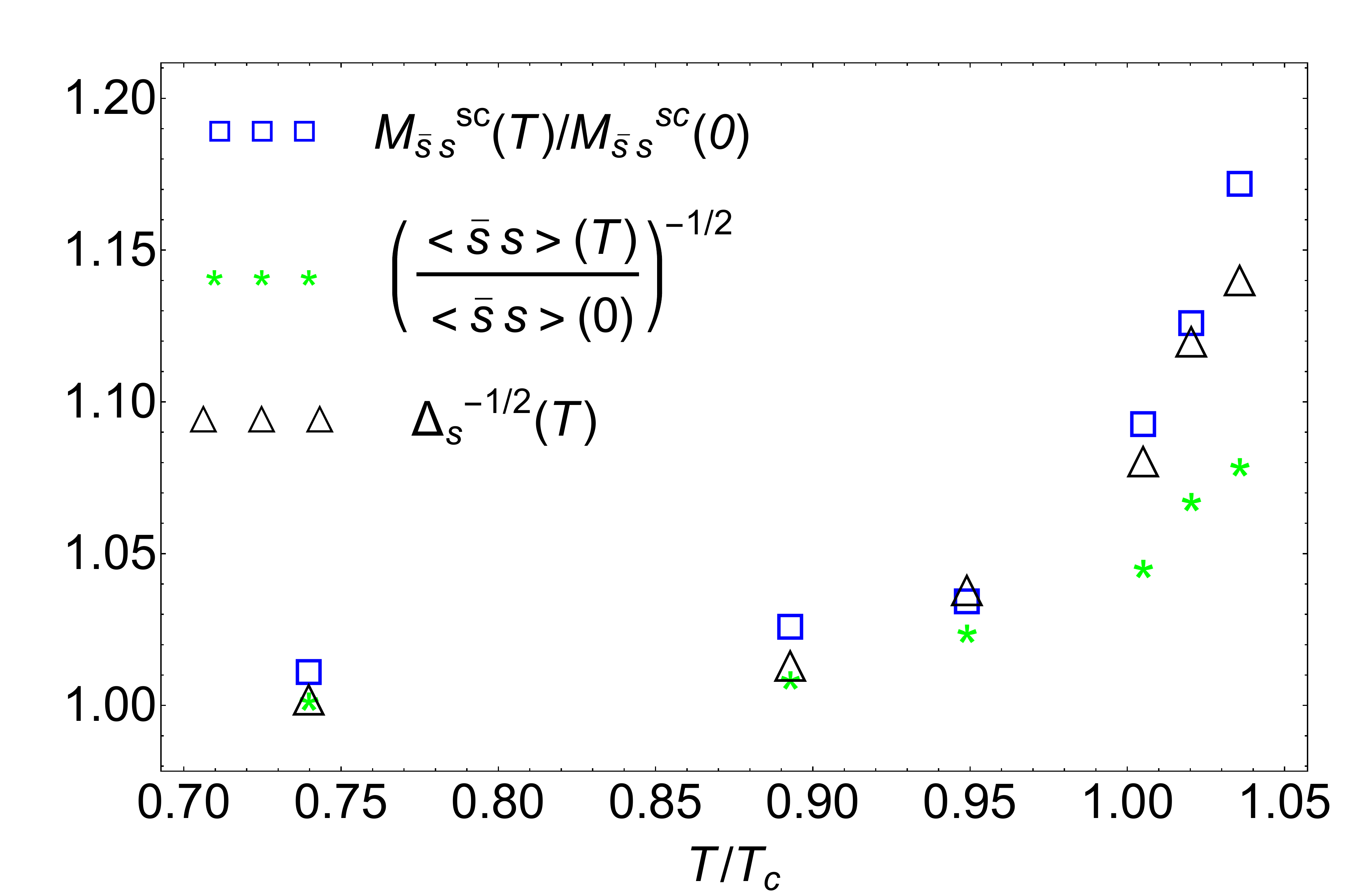}}
\caption{Comparison between the pseudoscalar screening mass ratio in the $\bar s s$ channel and the strange condensate ratios, subtracted and unsubtracted, defined in the main text. The  lattice data are taken from 
\cite{Cheng:2010fe} (masses) and \cite{Cheng:2007jq} (condensates) with the same lattice action and resolution and $T_c\simeq 196$ MeV. The values for $\Delta_s(T)$ correspond to the definition \eqref{Deltas} with $r_1^3\conds^{ref}=1.109$ and the value of $r_1\simeq$ 0.31 fm used in \cite{Cheng:2007jq}}
\label{fig:corrs}
\end{figure}
\end{center}

We show the results for these two channels in Figs.~\ref{fig:corrkaon}~and~\ref{fig:corrs}. We have set $r_1^3\conds^{ref}=1.109$ and, as before, $r_1^3\condl^{ref}=$ 0.749, which are the values minimizing the sum of the three channel squared differences between screening masses ratios and subtracted condensates $\Delta_{l,k,s}^{-1/2}$. In physical units $\conds^{ref}\simeq$ (637.47 MeV)$^3$, smaller than the average values for twice the strange condensate in \cite{Cheng:2007jq}, which are about (730 MeV)$^3$. The relative differences between the screening mass ratio and $\Delta_{K}^{-1/2}$ in Fig.~\ref{fig:corrkaon}  are below 3\%, while the differences with the direct  unsubtracted condensate ratios are about 11\% for the last point and below 6\% for the rest. As for the $\bar ss$-channel, the differences in Fig.~\ref{fig:corrs} between the screening mass ratio and $\Delta_{K}^{-1/2}$ are also below 3\%, 
whereas the difference with the unsubtracted values are about 8\% for the last point and below 5\% for the others. 

In addition, following the procedure described in \cite{Bazavov:2011nk}, we also show the results of leaving $\condl^{ref}=\conds^{ref}$ as the only free parameter of the fit. Doing so, we obtain $r_1^3\condl^{ref}=0.776$. The corresponding points are showed in Fig.~\ref{fig:corrall} for the three channels, where we also display the results of the previous fit with two free parameters for comparison. 
The relative deviations in the one-parameter case are below 7\%, 10\% and 6\% in the pion, $K$ and $\bar s s$ channel respectively.  
\begin{figure}
\centerline{\includegraphics[width=10cm]{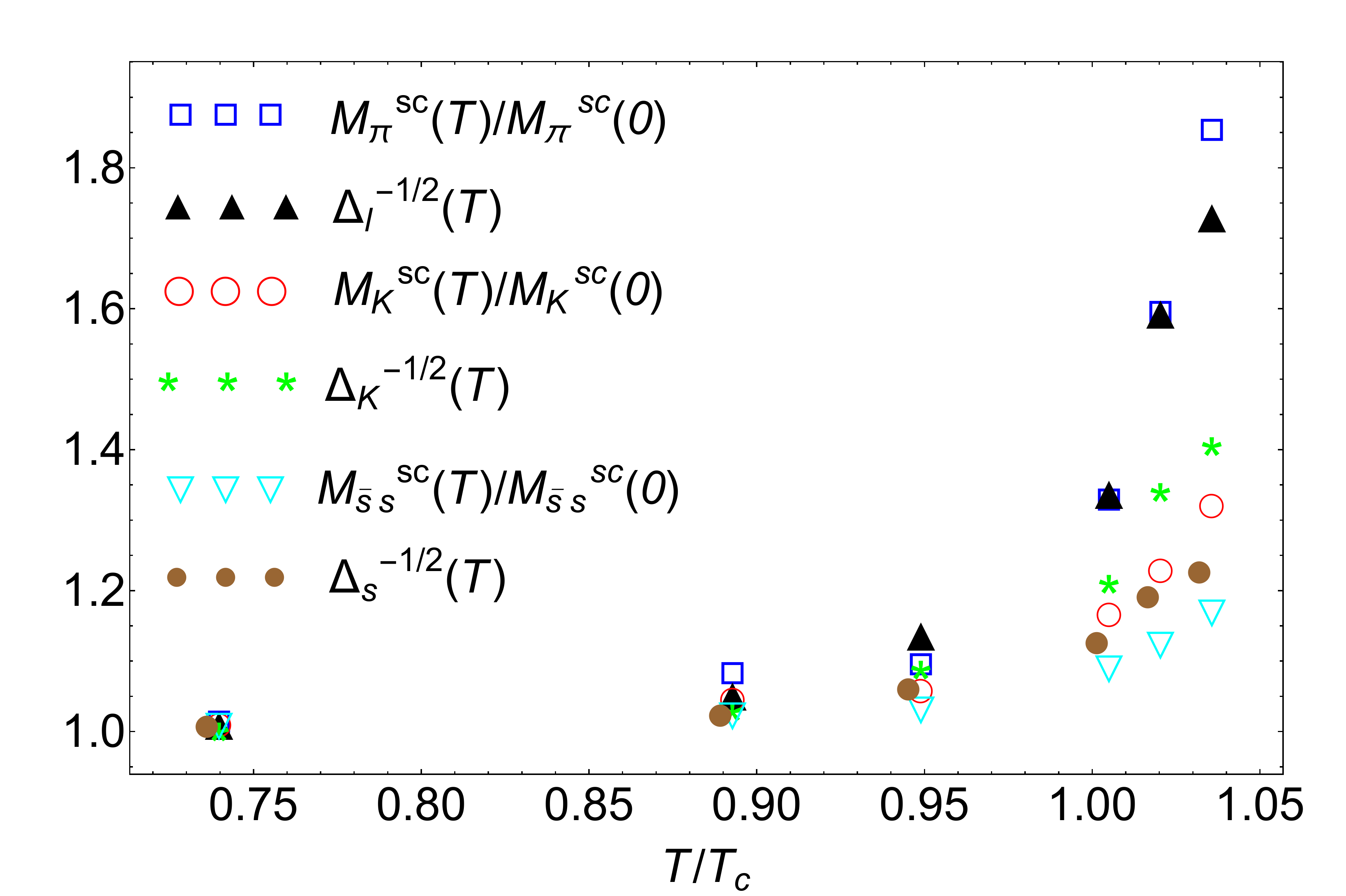}
\includegraphics[width=10cm]{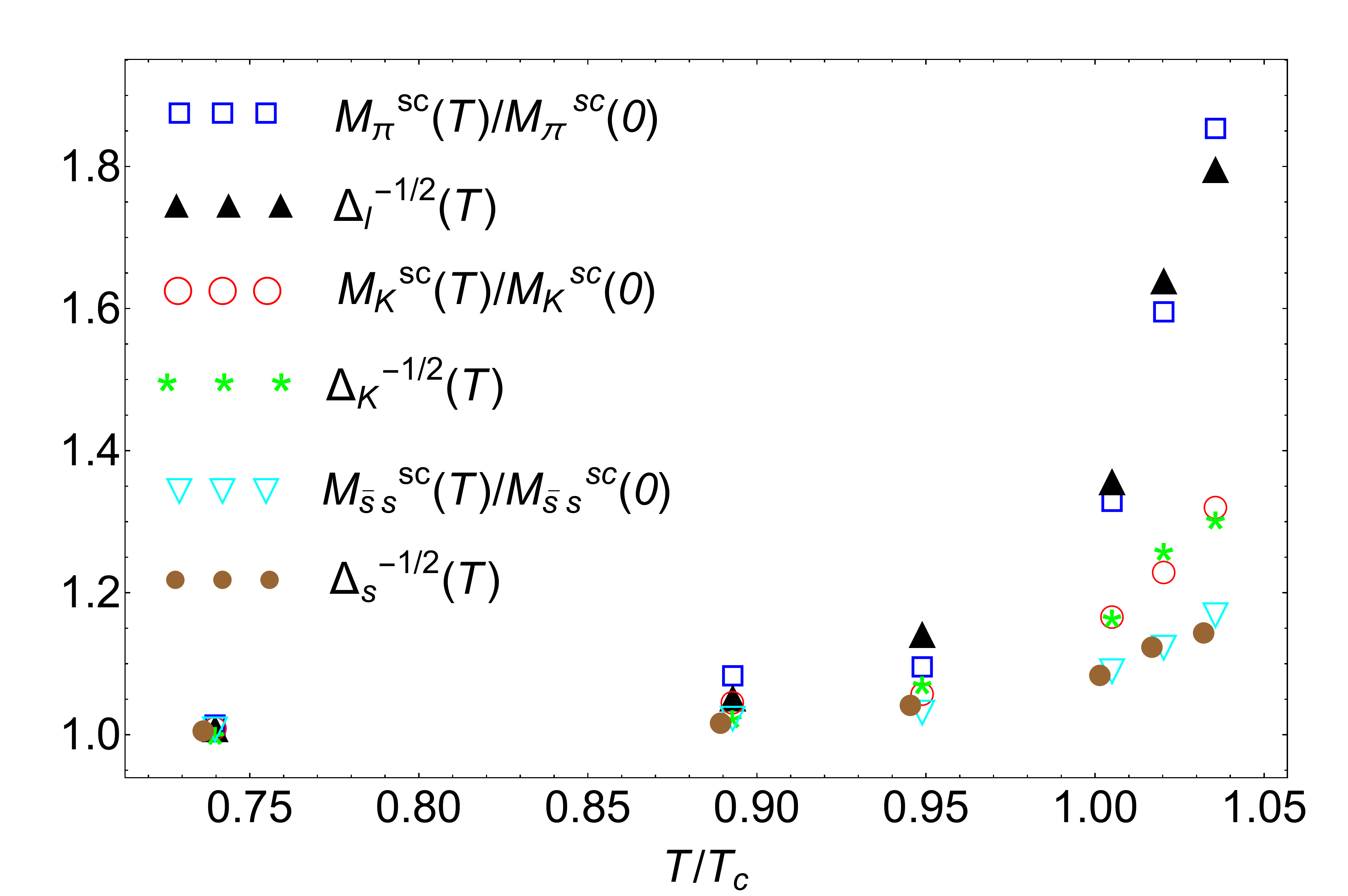}}
\caption{Comparison of pseudoscalar screening mass ratios and subtracted condensates for the three channels with reference values $r_1^3\condl^{ref}=r_1^3\conds^{ref}$=0.776 (left) and $r_1^3\condl^{ref}=0.749$ $r_1^3\conds^{ref}$=1.109 (right).  The  lattice data are taken from 
\cite{Cheng:2010fe} (masses) and \cite{Cheng:2007jq} (condensates) with the same lattice action and resolution, $T_c\simeq 196$ MeV. The values for $\Delta_l(T)$, $\Delta_K(T)$ and $\Delta_s(T)$ correspond respectively to the definitions \eqref{Deltal}, \eqref{DeltaK} and \eqref{Deltas} with and the value of $r_1\simeq$ 0.31 fm used in \cite{Cheng:2007jq}.}
\label{fig:corrall}
\end{figure}

Finally, we also explore the strange scaling law~\eqref{scalings} with the newer data for $\bar s s $ screening masses in \cite{Maezawa:2013nxa} and the corresponding condensate data in \cite{Bazavov:2011nk}, both with the same lattice conditions. Namely, a HISQ action, $N_\tau$=12 and $m_s=20\hat m$. The results for the unsubtracted and subtracted condensates are showed in Fig.~\ref{fig:corrsnew}.  As in~\cite{Bazavov:2011nk},  
we have taken $r_1^3\,\conds^{ref}=0.166$, which corresponds to the chiral limit $T=0$ condensate in \cite{Bazavov:2009bb}, i.e, we do not fit it to the squared differences. 
Even so, we obtain relative deviations below 4\%, which highlights again the importance of using proper subtracted condensates in the lattice. 
The differences with the unsubtracted condensate ratio are now around 13.3\% for the last point and less than 10\% for the others. 

\begin{center}
\begin{figure}
\centerline{\includegraphics[width=14cm]{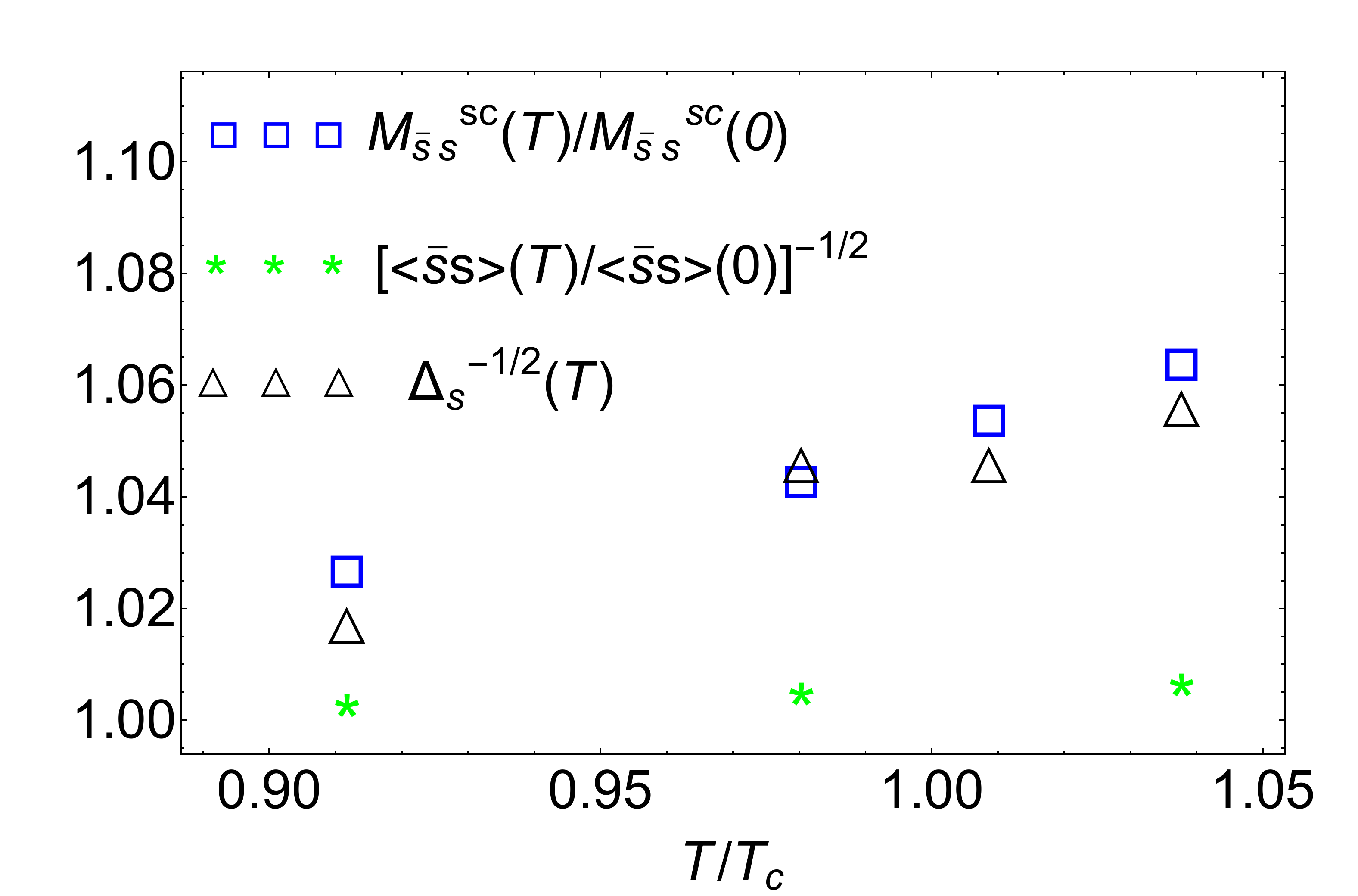}}
\caption{Comparison between the pseudoscalar screening mass ratio in the $\bar s s$ channel and the strange condensate ratios, subtracted and unsubtracted, defined in the main text. The  lattice data are taken from 
\cite{Maezawa:2013nxa} (masses) and \cite{Bazavov:2011nk} (condensates) with the same lattice action and resolution and $T_c\simeq 154$ MeV. The values for $\Delta_s(T)$ correspond to the definition \eqref{Deltas} with $r_1^3\conds^{ref}=0.166$ and the value of $r_1\simeq$ 0.31 fm used in \cite{Bazavov:2011nk}.}
\label{fig:corrsnew}
\end{figure}
\end{center}
Summarizing the results in this section, we observe a clear correlation of lattice screening masses and properly subtracted lattice condensates, which obey the scaling laws predicted by our results in \eqref{scalingpi}, \eqref{scalingK}, \eqref{scalings} with less than 5\% deviations, becoming higher for unsubtracted condensates due to the presence of lattice divergences. Recall also that in most of the cases analyzed, the largest deviations are around $(1-1.05) T_c$ where we are possibly surpassing  the applicability of our assumptions.

 \section{Conclusions}
 \label{sec:conc}
 
 In this work we have explored the relation between QCD quark condensates and pseudoscalar susceptibilities in the light meson sector with three light flavors, as well as  their  phenomenological consequences in connection with lattice data on meson screening masses. 
 
 We have derived formally the QCD Ward identities relating pseudoscalar susceptibilities and quark condensates for the pion, kaon and $\eta-\eta'$ channels, including the anomalous correlators entering for $U_A(1)$ transformations. In order to verify those identities, we have evaluated them in their low-energy representation provided by $SU(3)$ and $U(3)$ Chiral Perturbation Theory. The latter formalism is needed to incorporate consistently the $\eta'$ meson within the joint chiral and $1/N_c$ counting. Within this formalism, we have showed that the identities hold up to NLO in the chiral counting in $SU(3)$ and up to NNLO in the $\delta$-expansion in $U(3)$ ChPT. This is the first order in both formalisms at which temperature corrections enter through the meson loops. The full set of Ward identities for three flavors, the $U(3)$ ChPT quark condensates, as well as the $SU(3)$ and $U(3)$ pseudoscalar susceptibilities in the $\pi,K$ and $\eta-\eta'$ sectors, are new results of the present work not given elsewhere. 
  
 The second part of our analysis has dealt with the consequences of these identities for the behavior of the screening masses in the pion, kaon and $\bar s s $ channels as observed in lattice analysis. Assuming a soft temperature behavior for the pole-screening mass difference, as well as for the residues of the pseudoscalar correlators, these identities  predict a temperature scaling of masses related directly to that of quark condensates. Thus, the chiral restoring behavior of the light condensate induces a strong growing of the pion screening mass near the transition temperature, while the screening kaon mass has a softer behavior due to the contribution of the strange condensate. In the $\bar s s$ channel, the light condensate contribution cancels and the anomaly term is suppressed, so that the scaling is dominated by the strange condensate, producing  an even softer behavior. We have analyzed these scaling laws quantitatively, through a detailed comparison of  lattice results for screening masses and for condensates. We have also shown that it is particularly important to choose properly  lattice subtracted condensates  which behave as the continuum condensates, in order that they follow the mentioned scaling laws, thus avoiding quadratic lattice divergences. 
 
 We believe that the present analysis will be helpful to clarify several issues related to chiral symmetry restoration. First,  our  explicit derivation of all the pseudoscalar-condensate Ward identities involved for three light flavors provides a guideline for future lattice analysis. For instance the kaon channel identity has not been tested directly.  Second, our $SU(3)$ and $U(3)$ ChPT analysis  gives theoretical support to those identities, opening up also new possibilities, like the study of the $U_A(1)$ restoration, which  is also a topic of increasing interest  in recent lattice analysis \cite{Buchoff:2013nra,Bhattacharya:2014ara,Cossu:2015lnb}. Finally, our scaling law analysis for screening masses helps to understand in a very natural way the temperature behavior in the different channels in connection with chiral restoration.

 \section*{Acknowledgments}
We thank Z. H. Guo for helpful discussions.  Work partially supported by  research contracts FPA2011-27853-C02-02 (spanish ``Ministerio de Ciencia e Innovaci\'on"), FPA2014-53375-C2-2-P (spanish ``Ministerio de Econom\'{\i}a y Competitividad"). We also acknowledge the support
of the EU FP7 HadronPhysics3 project, the Spanish Hadron Excellence Network (spanish ``Ministerio de Econom\'{\i}a y Competitividad" contract FIS2014-57026-REDT), the DFG (SFB/TR 16, ``Subnuclear Structure of Matter'') and the UCM-Banco de Santander contract GR3/14 910309.
 
 \appendix
 \section{Detailed ChPT expressions}
 \label{sec:app}

Here we provide  explicit results obtained for the light and strange condensates, as well as for pseudoscalar susceptibilities within $U(3)$ ChPT, and mentioned in the main text. 

For the quark condensates at NNLO in the $\delta$ counting one gets in $U(3)$ ChPT:

\begin{align}
%-\left\langle\bar uu+\bar dd\right\rangle=
\condl(T)=&-2B^r_0F^2\left\{1+\frac{4}{F^2}\left[\left(4 L_6^r + 2 L_8^r+H_2^r \right) M_{0\pi}^2+8 L_6^r M_{0K}^2+\frac{12}{F^2} C^r_{19}M_{0\pi}^4\right]-3\mu_\pi (T) -2\mu_K (T)-\right.\nonumber\\
&\frac{1}{3}\left(\cth^2-2\sqrt 2\cth \sth+2\sth^2\right)\mu_\eta (T)-\frac{1}{3}\left(2\cth^2+2\sqrt 2 \cth \sth+\sth^2\right)\mu_{\eta^\prime}(T)-\nonumber\\
&\left.\frac{\cth\sth^2M_0^2\left[3\cth \sth M_0^2+2\left(\sqrt 2\cth^2-\cth \sth -\sqrt 2\sth^2\right)(M_{0K}^2-M_{0\pi}^2)\right]\left(\mu_\eta (T)-\mu_{\eta^\prime}(T)\right)}{(M_{0K}^2-M_{0\pi}^2)\left[3\sth M_0^2+2\left(2\sqrt 2\cth-\sth\right)
(M_{0K}^2-M_{0\pi}^2)\right]}\right\},\label{lightchptU3}\\
%-\left\langle\bar ss\right\rangle=
\conds(T)=&-B^r_0F^2\left\{1+\frac{4}{F^2}\left[\left(4 L_6^r - 2 L_8^r-H_2^r\right) M_{0\pi}^2+2\left(4 L_6^r + 2 L_8^r+H_2^r\right) M_{0K}^2 +\frac{12}{F^2}C_{19}^r\left(4M_{0K}^4-4M_{0\pi}^24M_{0K}^2+M_{0\pi}^4\right)\right]\right.\nonumber\\
&-4\mu_K (T)-\frac{2}{3}\left(2\cth^2+2\sqrt 2\cth \sth+\sth^2\right)\mu_{\eta} (T)-\frac{2}{3}\left(\cth^2-2\sqrt 2\cth \sth+2\sth^2\right)\mu_{\eta^\prime} (T)+\nonumber\\
&\left. \frac{2\cth\sth^2M_0^2\left[3\cth \sth M_0^2+2\left(\sqrt 2\cth^2-\cth \sth -\sqrt 2\sth^2\right)(M_{0K}^2-M_{0\pi}^2)\right]\left(\mu_\eta (T)-\mu_{\eta^\prime}(T)\right)}{(M_{0K}^2-M_{0\pi}^2)\left[3\sth M_0^2+2\left(2\sqrt 2\cth-\sth\right)
(M_{0K}^2-M_{0\pi}^2)\right]}\right\},\
\label{strangechptU3}
\end{align}
where $\cth\equiv\cos\theta$ and $\sth\equiv\sin\theta$, $\theta$ is  the $\eta-\eta^\prime$ mixing angle (not to be confused with the $\theta(x)$ field), $M_0^2$  is the anomalous part of the $\eta$ mass, which multiplies $X^2$ in the lagrangian ${\cal L}_{\delta^0}$ with $X=\log\det U + i\theta(x)$ and  $U$ the NGB matrix field, and the $\mu_i(T)$ are defined in \eqref{mudef}.   We follow the notation for the LECs in \cite{Guo:2015xva}, where the explicit expressions for the tree level $M_{0\eta}$ and $M_{0\eta^\prime}$ and for $\sth$ can also be found. The renormalization conditions for the LECs $L_i$, $C_i$ and $B_0$ are given in  \cite{Kaiser:2000gs}. 

As for the pseudoscalar susceptibilities in $U(3)$ for the $\eta-\eta^\prime$ sector, since the expressions are rather long we only provide explicitly here our result for $\chi_P^{8A}$, whereas the explicit expressions for the other susceptibilities involved can be obtained from the light and strange condensates and $\chi_P^{8A}$ through the identities \eqref{finaleqseta}, once we have verified that they hold as explained in the main text. In this expression, two additional LECs enter, namely $\Lambda_2$ and $v_2^{(2)}$ in the notation of \cite{Guo:2015xva}.  We get:
 
\allowdisplaybreaks[3]
\begin{align}
\chi_{P,LO}^{8A}=&-\frac{2\sqrt 6 B^r_0F^2c_\theta s_\theta}{\meta^2 \metap^2}\left(\metap^2-\meta^2\right),\nonumber\\
\chi_{P,NLO}^{8A}=&\chi_{P,LO}^{8A}+\frac{2\sqrt 6 B^r_0F^2}{M_{0\eta}^2 M_{0\eta^\prime}^2}\Bigg\{\frac{\Lambda_2}{3}\Bigg[M_0^2 c_{2\theta}\left(2\left(\sqrt2c_{2\theta}+s_{2\theta}\right)M_{0K}^2-\left(2\sqrt 2c_{2\theta}-s_{2\theta}\right)M_{0\pi}^2\right)\nonumber\\
&-\cth\left(1-\frac{2M_0^2\sth^2}{\meta^2}\right)\left(2\left(\sqrt 2\cth+\sth\right)\mk^2-\left(2\sqrt2\cth-\sth\right)\mpi^2\right)\metap^2\nonumber\\
&+\left.\!\!\sth\left(1-\frac{2M_0^2\cth^2}{\metap^2}\right)\left(2\left(\cth-\sqrt2\sth\right)\mk^2+\left(\cth+2\sqrt2\sth\right)\mpi^2\right)\meta^2 \right]\nonumber\\
&+\frac{16M_0^2L_{8}^r}{3F^2}\Bigg[2c_{2\theta}\left(2\sqrt2c_{2\theta}-s_{2\theta}\right)\mk^2\left(\mk^2-\mpi^2\right)\nonumber\\
&-\sth\left(2\sqrt2\left(\sqrt2\cth+\sth\right)\mk^2-\left(\cth+2\sqrt2\sth\right)\mpi^2\right.
\nonumber\\
&-\left.\frac{\cth}{\meta^2}\left(3\mpi^4+4\left(\sqrt2\cth+\sth\right)^2\mk^2\left(\mk^2-\mpi^2\right)\right)\right)\metap^2\nonumber\\ 
&-\cth\left(2\sqrt2\left(\cth-\sqrt2\sth\right)\mk^2-\left(2\sqrt2\cth-\sth\right)\mpi^2\right.
\nonumber\\
&+\left.\frac{\sth}{\metap^2}\left(3\mpi^4+4\left(\cth-\sqrt2\sth\right)^2\mk^2\left(\mk^2-\mpi^2\right)\right)\right)\meta^2\Bigg]\Bigg\}\nonumber\\
\chi_{P,NNLO}^{8A}=&\chi_{P,NLO}^{8A}-\frac{\sqrt 3  B^r_0 F^2}{\meta^6 \metap^6}\Bigg\{\frac{\Lambda _2^2}{36} \Bigg[-8 M_0^2 c_{6\theta} \left(-\mk^2 \mpi^2+2 \mk^4-\mpi^4\right) \left(\meta^2-\metap^2\right){}^3\nonumber\\
&-16 c_{4\theta} \left(-\mk^2 \mpi^2+2 \mk^4-\mpi^4\right) \left(\meta^2-\metap^2\right){}^2\left(\meta^2 \left(M_0^2-\metap^2\right)+M_0^2 \metap^2\right)\nonumber\\
&+8 M_0^2 c_{2\theta} \left(-\mk^2 \mpi^2+2\mk^4-\mpi^4\right) \left(\meta^2+\metap^2\right){}^2 \left(\meta^2-\metap^2\right)\nonumber\\
&+16 \left(-\mk^2\mpi^2+2 \mk^4-\mpi^4\right) \left(\meta^4+\metap^4\right) \left(\meta^2 \left(M_0^2-\metap^2\right)+M_0^2 \metap^2\right)\nonumber\\
&+\sqrt{2} M_0^2 s_{6\theta} \left(-20 \mk^2 \mpi^2+4 \mk^4+7 \mpi^4\right)\left(\meta^2-\metap^2\right){}^3\nonumber\\
&-2 \sqrt{2} s_{4\theta} \left(\meta^2-\metap^2\right){}^2 \bigg(4 \mk^2 \mpi^2 \left(\meta^2 \left(2 M_0^2-5 \metap^2\right)+2 M_0^2 \metap^2\right)\nonumber\\
&+4 \mk^4 \left(\meta^2 \left(\metap^2+2 M_0^2\right)+2 M_0^2 \metap^2\right)+\mpi^4 \left(\meta^2 \left(7 \metap^2+2 M_0^2\right)+2 M_0^2 \metap^2\right)\bigg)\nonumber\\
&-\sqrt{2} s_{2\theta} \left(\meta^2-\metap^2\right) \bigg(-4 \mk^2 \mpi^2 \left(\meta^4 \left(7M_0^2-12 \metap^2\right)+6 \meta^2 \left(3 M_0^2 \metap^2-2 \metap^4\right)+7 M_0^2 \metap^4\right)\nonumber\\
&+\mk^4\left(\meta^4 \left(44 M_0^2-48 \metap^2\right)+24 \meta^2 \left(3 M_0^2 \metap^2-2 \metap^4\right)+44 M_0^2\metap^4\right)\nonumber\\
&+\mpi^4 \left(\meta^4 \left(29 M_0^2-36 \metap^2\right)+18 \meta^2 \left(3 M_0^2 \metap^2-2\metap^4\right)+29 M_0^2 \metap^4\right)\bigg)\Bigg]\nonumber\\
&+6 \sqrt{2} v_2^{(2)} \meta^2 \metap^2 s_{2\theta} \left(2 \mk^2+\mpi^2\right) \left(\meta^2-\metap^2\right)\left(\meta^2 \left(M_0^2 c_{\theta}^2-\metap^2\right)+M_0^2 \metap^2 s_{\theta}^2\right)\nonumber\\
&+\frac{4 L_{25}^r \metap^2}{F^2}\Bigg[-8 c_{2\theta} \meta^4 \metap^2 \left(M_0^2-2 \mk^2\right) \left(\mk^2-\mpi^2\right) \left(\meta^2-\metap^2\right)\nonumber\\
&-8 M_0^2 c_{4\theta} \mk^2 \left(\mk^2-\mpi^2\right) \left(\meta^3-\meta \metap^2\right){}^2\nonumber\\
&+8 \meta^2 \left(\mk^2-\mpi^2\right) \left(\mk^2 \left(\meta^2+\metap^2\right)-\meta^2 \metap^2\right) \left(\meta^2 \left(M_0^2-2 \metap^2\right)+M_0^2 \metap^2\right)\nonumber\\
&-2 \sqrt{2} \meta^2 s_{2\theta} \left(-4 \mk^2 \mpi^2+4 \mk^4+3 \mpi^4\right) \left(\meta^2-\metap^2\right) \left(\meta^2 \left(M_0^2-\metap^2\right)+M_0^2 \metap^2\right)\nonumber\\
&-\sqrt{2} M_0^2 s_{4\theta} \left(-4 \mk^2 \mpi^2+4 \mk^4+3 \mpi^4\right) \left(\meta^3-\meta \metap^2\right){}^2\Bigg]\nonumber\\
&+\frac{8L_6^r M_0^2 \metap^2}{3F^2} \Bigg[4 c_{4\theta} \left(-\mk^2 \mpi^2+2 \mk^4-\mpi^4\right) \left(\meta^3-\meta \metap^2\right){}^2\nonumber\\
&-4 \left(-\mk^2 \mpi^2+2 \mk^4-\mpi^4\right) \left(\meta\metap^2+\meta^3\right){}^2\nonumber\\
&+6 \sqrt{2}\meta^2 s_{2\theta} \left(2 \mk^2+\mpi^2\right) \left(\meta^2-\metap^2\right) \left(\mk^2 \left(\meta^2+\metap^2\right)-\meta^2 \metap^2\right)\nonumber\\
&-\sqrt{2} s_{4\theta}\left(\mk^2-\mpi^2\right) \left(2 \mk^2+\mpi^2\right) \left(\meta^3-\meta \metap^2\right){}^2\Bigg]\nonumber\\
&+\frac{4 L_7^r M_0^2 \metap^2}{3 F^2} \Bigg[8 c_{4\theta} \left(-\mk^2 \mpi^2+2 \mk^4-\mpi^4\right) \left(\meta^3-\meta \metap^2\right){}^2\nonumber\\
&-8 \left(-\mk^2 \mpi^2+2 \mk^4-\mpi^4\right) \left(\meta\metap^2+\meta^3\right){}^2
\nonumber\\
&+6 \sqrt{2} \meta^2 s_{2\theta} \left(-4 \mk^2 \mpi^2+4 \mk^4+3 \mpi^4\right) \left(\meta^4-\metap^4\right)\nonumber\\
&-\sqrt{2} s_{4\theta} \left(-20 \mk^2 \mpi^2+4 \mk^4+7 \mpi^4\right)\left(\meta^3-\meta \metap^2\right){}^2\Bigg]\nonumber\\
&+\frac{8L^r_8\Lambda _2} {9 F^2}\Bigg[-8 c_{6\theta} \mk^2 \left(\mk^4+\mpi^2 \mk^2-2 \mpi^4\right) M_0^2 \left(\meta^2-\metap^2\right){}^3\nonumber\\
&+2 \sqrt{2} \mk^2 M_0^2 s_{6\theta}\left(10 \mk^4-17 \mpi^2 \mk^2+7 \mpi^4\right) \left(\meta^2-\metap^2\right){}^3\nonumber\\
&+8 c_{4\theta} \left(\mk^2-\mpi^2\right) \bigg(\left(\meta^2 \left(\metap^2-10 M_0^2\right)-10 \metap^2 M_0^2\right) \mk^4\nonumber\\
&+\left(\left(2 \left(\metap^2+2 M_0^2\right) \mpi^2+\metap^2M_0^2\right) \meta^2+4 \metap^2 \mpi^2 M_0^2\right) \mk^2\nonumber\\
&+\mpi^2 \left(-\meta^2 \metap^2-3\left(\meta^2+\metap^2\right) \mpi^2\right) M_0^2\bigg) \left(\meta^2-\metap^2\right){}^2\nonumber\\
&-\sqrt{2}s_{4\theta} \left(\meta^2-\metap^2\right){}^2\bigg(4 \left(\left(5\metap^2+4 M_0^2\right) \meta^2+4 \metap^2 M_0^2\right) \mk^6\nonumber\\
&+2 \left(-\left(\left(17 \metap^2+4 M_0^2\right)\mpi^2+8 \metap^2 M_0^2\right) \meta^2-4 \metap^2 \mpi^2 M_0^2\right) \mk^4 \nonumber\\
&+2 \mpi^2 \left(\left(7\left(\mpi^2+M_0^2\right) \metap^2+2 \mpi^2 M_0^2\right) \meta^2+2 \metap^2 \mpi^2 M_0^2\right)\mk^2\nonumber\\
&+\mpi^4 \left(6 \left(\meta^2+\metap^2\right) \mpi^2-7 \meta^2 \metap^2\right) M_0^2\bigg)\nonumber\\
&+8 c_{2\theta} \left(\mk^2-\mpi^2\right) \bigg(\left(\meta^2+\metap^2\right) \left(\left(8 \metap^2+M_0^2\right) \meta^2+\metap^2 M_0^2\right) \mk^4\nonumber\\
&+2 \left(-\left(\metap^4+2 M_0^2\metap^2\right) \meta^4-2 \metap^4 M_0^2 \meta^2+\left(\meta^2+\metap^2\right) \mpi^2\left(\left(M_0^2-4 \metap^2\right) \meta^2+\metap^2 M_0^2\right)\right) \mk^2\nonumber\\
&+\meta^2 \metap^2\mpi^2 \left(\left(2 \metap^2+3 \mpi^2-2 M_0^2\right) \meta^2+\metap^2 \left(3 \mpi^2-2 M_0^2\right)\right)\bigg)\left(\meta^2-\metap^2\right)\nonumber\\
&-2 \sqrt{2}s_{2\theta} \left(\meta^2-\metap^2\right) \bigg\{\left(\left(46 M_0^2-28 \metap^2\right) \meta^4+4 \left(17 \metap^2 M_0^2-7\metap^4\right) \meta^2+46 \metap^4 M_0^2\right) \mk^6\nonumber\\
&+\bigg(16 \metap^2 \left(\metap^2-M_0^2\right)\meta^4-16 \metap^4 M_0^2 \meta^2\nonumber\\
&+\mpi^2 \left(\left(38 \metap^2-59 M_0^2\right) \meta^4+\left(38\metap^4-106 \metap^2 M_0^2\right) \meta^2-59 \metap^4 M_0^2\right)\bigg) \mk^4\nonumber\\
&+\mpi^2 \bigg(14 \meta^2 \left(\left(M_0^2-\metap^2\right) \meta^2+\metap^2 M_0^2\right) \metap^2\nonumber\\
&+\left(\meta^2+\metap^2\right) \mpi^2 \left(\left(25 M_0^2-16 \metap^2\right) \meta^2+25 \metap^2 M_0^2\right)\bigg) \mk^2\nonumber\\
&-7 \meta^2 \metap^2 \mpi^4 \left(\left(M_0^2-\metap^2\right) \meta^2+\metap^2 M_0^2\right)\nonumber\\
&+3 \mpi^6\left(\left(2 M_0^2-\metap^2\right) \meta^4+\left(2 \metap^2 M_0^2-\metap^4\right) \meta^2+2 \metap^4M_0^2\right)\bigg\} \nonumber\\
&+8 \left(\mk^2-\mpi^2\right) \bigg\{3\left(\meta^4+\metap^4\right) \left(\left(M_0^2-\metap^2\right) \meta^2+\metap^2 M_0^2\right) \mpi^4\nonumber\\
&-\left(\meta^4+\metap^4\right) \left(\left(\left(M_0^2-6 \mk^2\right) \meta^2+4 \mk^2 M_0^2\right) \metap^2+4 \mk^2\meta^2 M_0^2\right) \mpi^2\nonumber\\
&+\mk^2 \bigg(10 \mk^2 M_0^2 \meta^6+\metap^2 \left(\mk^2 \left(10 M_0^2-9\meta^2\right)-5 \meta^2 M_0^2\right) \meta^4\nonumber\\
&+2 \metap^4 \left(3 \meta^4-3 \left(\mk^2+M_0^2\right)\meta^2+5 \mk^2 M_0^2\right) \meta^2\nonumber\\
&+\metap^6 \left(6 \meta^4-5 M_0^2 \meta^2+\mk^2 \left(10 M_0^2-9\meta^2\right)\right)\bigg)\bigg\}\Bigg]\nonumber\\
&+ \frac{8  M_0^2 \metap^2}{3 f^4}\left(3 C_{19}+2 C_{31}\right) \Bigg[16 c_{2\theta} \mk^2 \meta^4 \metap^2 \left(\mk^2-\mpi^2\right) \left(\meta^2-\metap^2\right)\nonumber\\
&+4 c_{4\theta} \left(-6 \mk^4 \mpi^2+3 \mk^2 \mpi^4+4\mk^6-\mpi^6\right) \left(\meta^3-\meta \metap^2\right){}^2\nonumber\\
&-2 \sqrt{2} \meta^2 s_{2\theta} \left(\meta^2-\metap^2\right) \bigg(2 \mk^4 \left(9 \mpi^2\left(\meta^2+\metap^2\right)+4 \meta^2 \metap^2\right)\nonumber\\
&+\mk^2 \mpi^2 \left(-9 \mpi^2\left(\meta^2+\metap^2\right)-8 \meta^2 \metap^2\right)-12 \mk^6 \left(\meta^2+\metap^2\right)+3\meta^2 \metap^2 \mpi^4\bigg)\nonumber\\
&-\sqrt{2} s_{4\theta} \left(-6 \mk^4 \mpi^2+3 \mk^2 \mpi^4+4\mk^6-\mpi^6\right) \left(\meta^3-\meta \metap^2\right){}^2\nonumber\\
&-4 \meta^2 \left(\mk^2-\mpi^2\right) \left(\meta^2+\metap^2\right) \bigg(-2 \mk^2 \left(\mpi^2 \left(\meta^2+\metap^2\right)+2\meta^2 \metap^2\right)\nonumber\\
&+4 \mk^4 \left(\meta^2+\metap^2\right)+\mpi^4 \left(\meta^2+\metap^2\right)\bigg)\Bigg]\nonumber\\
&+\frac{\mu_\pi (T) M_0^2\meta^2\metap^2\mpi^2}{2F^2}\Bigg[c_{\theta}\meta^4\left(4\metap^2\left(2c_{\theta}+\sqrt{2}s_{\theta}\right)-2 \mpi^2s_{\theta}\left(\sqrt{2}c_{2\theta}+4s_{2\theta}+3\sqrt{2}\right)\right)\nonumber\\
&+\meta^2\metap^2\left(\mpi^2\left(-4c_{4\theta}+\sqrt{2}s_{4\theta}-4\right)-4\metap^2s_{\theta}\left(\sqrt{2}c_{\theta}-2s_{\theta}\right)\right)-\metap^4\mpi^2s_{2\theta}\left(\sqrt{2}c_{2\theta}+4s_{2\theta}-3 \sqrt{2}\right)\Bigg]\nonumber\\
&+\frac{\mu_K (T) M_0^2\mk^2\meta^2\metap^2}{3 F^2}\Bigg[\meta^2\metap^2\left(4\mk^2\left(2c_{4\theta}+\sqrt{2}s_{4\theta}+2\right)+8c_{\theta}\meta^2\left(\sqrt{2}s_{\theta}-c_{\theta}\right)\right.
\nonumber\\
&+\left.\mpi^2\left(-4c_{4\theta}+\sqrt{2}s_{4\theta}-4\right)\right)
%\nonumber\\
%&
-4\metap^4s_{\theta}\left(-2\mk^2 s_{2\theta}\left(2c_{\theta}+\sqrt{2}s_{\theta}\right)+2\meta^2\left(\sqrt{2}c_{\theta}+s_{\theta}\right)\right.
\nonumber\\
&+\left.c_{\theta}^2 \mpi^2 \left(\sqrt{2} c_{\theta}+4 s_{\theta}\right)\right)
%\nonumber\\
%&
+4 c_{\theta} \meta^4 s_{\theta} \left(\mpi^2 s_{\theta}\left(\sqrt{2} s_{\theta}-4 c_{\theta}\right)-4 c_{\theta} \mk^2 \left(\sqrt{2} c_{\theta}-2 s_{\theta}\right)\right)\Bigg]\nonumber\\
&+\frac{M_0^2 \mu _{\eta}\meta^4 \metap^2 }{144 F^2}\Bigg[4 \mk^2 \bigg(4 c_{2\theta} \meta^2 \metap^2 \left(24 c_{2\theta}+8 c_{4\theta}+\sqrt{2} \left(7 s_{4\theta}-6 s_{2\theta}\right)\right)\nonumber\\
&-2\meta^4 s_{2\theta}\left(7 \sqrt{2} c_{4\theta}-8 s_{4\theta}+9 \sqrt{2}\right)+2 \metap^4 s_{2\theta} \left(12 \sqrt{2} c_{2\theta}-7 \sqrt{2} c_{4\theta}+48 s_{2\theta}+8 s_{4\theta}+27\sqrt{2}\right)\bigg)\nonumber\\
&-2\metap^4 s_{2\theta} \bigg(36\meta^2\left(-\sqrt{2} c_{2\theta}+2 s_{2\theta}+3 \sqrt{2}\right)+\mpi^2 \left(36\sqrt{2}c_{2\theta}-7 \sqrt{2} c_{4\theta}+144 s_{2\theta}+8 s_{4\theta}+27\sqrt{2}\right)\bigg)\nonumber\\
&-4\meta^2\metap^2\left(6\meta^2\left(c_{\theta}+3c_{3\theta}\right)\left(2c_{\theta}+\sqrt{2}s_{\theta}\right)+c_{2\theta}\mpi^2\left(72c_{2\theta}+8c_{4\theta}+\sqrt{2}\left(7s_{4\theta}-18 s_{2\theta}\right)\right)\right)\nonumber\\
&+2 \meta^4\mpi^2 s_{2\theta} \left(7 \sqrt{2} c_{4\theta}-8 s_{4\theta}+9 \sqrt{2}\right)\Bigg]\nonumber\\
&\frac{M_0^2 \mu_{\eta^\prime} (T) \meta^2 \metap^4}{72 F^2} \Bigg[\meta^4 s_{2\theta} \bigg(4 \mk^2 \left(12 \sqrt{2} c_{2\theta}+7 \sqrt{2} c_{4\theta}+48 s_{2\theta}-8 s_{4\theta}-27 \sqrt{2}\right)\nonumber\\
&+36 \metap^2 \left(\sqrt{2} c_{2\theta}-2 s_{2\theta}+3 \sqrt{2}\right)+\mpi^2 \left(-36 \sqrt{2}c_{2\theta}-7 \sqrt{2} c_{4\theta}-144 s_{2\theta}+8 s_{4\theta}+27 \sqrt{2}\right)\bigg)\nonumber\\
&+2\meta^2\metap^2\bigg(c_{2\theta}\left(\mpi^2\left(-72c_{2\theta}+8c_{4\theta}+\sqrt{2}\left(18s_{2\theta}+7s_{4\theta}\right)\right)-4\mk^2\left(-24c_{2\theta}+8c_{4\theta}+\sqrt{2}\left(6s_{2\theta}+7s_{4\theta}\right)\right)\right)\nonumber\\
&+6 \metap^2 \left(\sqrt{2} c_{\theta}-2 s_{\theta}\right) \left(s_{\theta}-3 s_{3\theta}\right)\bigg)+\metap^4 s_{2\theta}\left(7 \sqrt{2} c_{4\theta}-8 s_{4\theta}+9 \sqrt{2}\right) \left(4 \mk^2-\mpi^2\right)\Bigg]\nonumber\\
&+\frac{256L^{r\,2}_8 M_0^2}{9 f^4} \Bigg[\bigg(8 c_{6\theta} \left(\meta^2-\metap^2\right){}^3+7\sqrt{2}s_{6\theta}\left(\meta^2-\metap^2\right){}^3-48c_{4\theta}\left(\meta^2+\metap^2\right)\left(\meta^2-\metap^2\right){}^2\nonumber\\
&+12 \sqrt{2} \left(\meta^2+\metap^2\right) s_{4\theta} \left(\meta^2-\metap^2\right){}^2-8 c_{2\theta} \left(\meta^2+\metap^2\right){}^2\left(\meta^2-\metap^2\right)\nonumber\\
&+48 \left(\meta^2+\metap^2\right) \left(\meta^4+\metap^4\right)+\sqrt{2}\left(-61 \meta^6-25 \metap^2 \meta^4+25 \metap^4 \meta^2+61 \metap^6\right) s_{2\theta}\bigg)\mk^8\nonumber\\
&+2\bigg(-8 c_{6\theta} \mpi^2 \left(\meta^2-\metap^2\right){}^3-7 \sqrt{2} \mpi^2 s_{6\theta} \left(\meta^2-\metap^2\right){}^3\nonumber\\
&+2 c_{4\theta} \left(5 \meta^2 \metap^2+24 \left(\meta^2+\metap^2\right) \mpi^2\right) \left(\meta^2-\metap^2\right){}^2\nonumber\\
&+2 \sqrt{2} s_{4 \theta}\left(\meta^2 \metap^2-6 \left(\meta^2+\metap^2\right) \mpi^2\right) \left(\meta^2-\metap^2\right){}^2\nonumber\\
&+\sqrt{2}s_{2\theta}\left(20 \meta^2 \left(\meta^2+\metap^2\right) \metap^2+\left(61 \meta^4+86 \metap^2 \meta^2+61 \metap^4\right) \mpi^2\right)\left(\meta^2-\metap^2\right)\nonumber\\
&-48 \left(\meta^2+\metap^2\right) \left(\meta^4+\metap^4\right) \mpi^2-6 \meta^2 \metap^2 \left(3 \meta^4+2 \metap^2 \meta^2+3 \metap^4\right)\nonumber\\
&+8 c_{2\theta} \left(\meta^4-\metap^4\right) \left(\left(\meta^2+\metap^2\right) \mpi^2-\meta^2 \metap^2\right)\bigg) \mk^6\nonumber\\
&+\mpi^2 \bigg(8 c_{6\theta} \mpi^2 \left(\meta^2-\metap^2\right){}^3+7\sqrt{2} \mpi^2 s_{6\theta} \left(\meta^2-\metap^2\right){}^3\nonumber\\
&-4 c_{4\theta} \left(7 \meta^2 \metap^2+18\left(\meta^2+\metap^2\right) \mpi^2\right) \left(\meta^2-\metap^2\right){}^2\nonumber\\
&+\sqrt{2}s_{4\theta}\left(18 \left(\meta^2+\metap^2\right) \mpi^2-11 \meta^2 \metap^2\right)\left(\meta^2-\metap^2\right){}^2\nonumber\\
&-\sqrt{2}s_{2\theta}\left(62 \meta^2 \left(\meta^2+\metap^2\right) \metap^2+\left(97 \meta^4+122 \metap^2 \meta^2+97 \metap^4\right) \mpi^2\right) \left(\meta^2-\metap^2\right)\nonumber\\
&+72 \left(\meta^2+\metap^2\right) \left(\meta^4+\metap^4\right) \mpi^2+12 \meta^2 \metap^2 \left(5 \meta^4+2 \metap^2 \meta^2+5 \metap^4\right)\nonumber\\
&-8 c_{2\theta} \left(\meta^4-\metap^4\right) \left(\left(\meta^2+\metap^2\right) \mpi^2-4 \meta^2 \metap^2\right)\bigg) \mk^4\nonumber\\
&+\mpi^4 \bigg(28 c_{2\theta}\meta^2 \left(\metap^4-\meta^4\right) \metap^2-12 \left(\meta^4+\metap^4\right) \left(3 \meta^2 \metap^2+2 \left(\meta^2+\metap^2\right) \mpi^2\right)\nonumber\\
&+8 c_{4\theta} \left(\meta^2-\metap^2\right){}^2 \left(\meta^2 \metap^2+3 \left(\meta^2+\metap^2\right) \mpi^2\right)\nonumber\\
&+2\sqrt{2}s_{2 \theta}\left(17 \meta^2 \left(\meta^4-\metap^4\right) \metap^2+18 \left(\meta^6-\metap^6\right) \mpi^2\right)\nonumber\\
&+\sqrt{2} \left(\meta^2-\metap^2\right){}^2 \left(7 \meta^2 \metap^2-6 \left(\meta^2+\metap^2\right)\mpi^2\right) s_{4\theta}\bigg) \mk^2\nonumber\\
&+6\meta^2\metap^2\mpi^6\left(c_{\theta}\left(4c_{\theta}-\sqrt{2}s_{\theta}\right)\meta^4+\metap^4s_{\theta}\left(\sqrt{2}c_{\theta}+4s_{\theta}\right)\right)+9\sqrt{2}s_{2\theta}\left(\metap^6-\meta^6\right)\mpi^8\Bigg]\Bigg\}
\label{chptchi8A} 
\end{align}

\end{document}